\documentclass[11pt]{article}

\usepackage[numbers,sort&compress]{natbib} 
\usepackage{amsmath}
\usepackage{graphicx}
\usepackage{amsfonts}
\usepackage{bm}
\usepackage{amssymb}
\usepackage{epsfig}
\usepackage{color}
\usepackage{psfrag}
\usepackage{mathrsfs}
\usepackage{esint}
\usepackage{pifont}
\usepackage{amsthm}
\usepackage{url}
\numberwithin{equation}{section}
\usepackage{stackengine}
\usepackage{comment}
\usepackage[mathscr]{euscript}
\usepackage[cal=esstix,scr=boondox,bb= pazo]{mathalfa}
\usepackage{stmaryrd,scalerel} 
\usepackage{subcaption}

\usepackage{tikz}
\usetikzlibrary{calc}
\usepackage{relsize}
\usepackage{bm}
\usepackage{moresize}
\usepgflibrary{shapes.geometric}
\usetikzlibrary{chains, positioning, quotes}
\usetikzlibrary{shapes,snakes}
\usetikzlibrary{snakes}
\usetikzlibrary{automata}
\usetikzlibrary{positioning}
\usetikzlibrary{arrows,decorations.pathmorphing,patterns,backgrounds,positioning,fit,petri}
\usetikzlibrary{automata}
\usetikzlibrary{graphs}
\usetikzlibrary{shadows}
\definecolor{colordiagram1}{RGB}{155, 255, 217}
\definecolor{colordiagram2}{RGB}{255, 148, 37} 
\definecolor{myyellow}{RGB}{255, 207, 0}
\definecolor{mygray}{RGB}{200, 200, 200}

\newcommand{\Os}{ {\color{black}{ {\mathcal O} }} }
\newcommand{\im}{{\rm i}}

\newcommand{\rd}{{\rm d}}

\def\dashint{\,\ThisStyle{\ensurestackMath{\stackinset{c}{.2\LMpt}{c}{.5\LMpt}{\SavedStyle-}{\SavedStyle\phantom{\int}}}\setbox0=\hbox{$\SavedStyle\int\,$}\kern-\wd0}\int}

\begin{document} \setcounter{page}{0}
\topmargin 0pt
\oddsidemargin 5mm
\renewcommand{\thefootnote}{\arabic{footnote}}
\newpage
\setcounter{page}{1}
\topmargin 0pt
\oddsidemargin 5mm
\renewcommand{\thefootnote}{\arabic{footnote}}
\newpage

\begin{titlepage}
\begin{flushright}
\end{flushright}
\vspace{0.5cm}
\begin{center}
{\large {\bf Droplet-mediated long-range interfacial correlations. Exact results for entropic repulsion effects.}}\\
\vspace{1.8cm}
{\large Alessio Squarcini$^{1,2,\natural}$ and Antonio Tinti$^{3,\flat}$ }\\
\vspace{0.5cm}
{\em $^1$Max-Planck-Institut f\"ur Intelligente Systeme,\\
Heisenbergstr. 3, D-70569, Stuttgart, Germany}\\
{\em $^2$IV. Institut f\"ur Theoretische Physik, Universit\"at Stuttgart,\\
Pfaffenwaldring 57, D-70569 
Stuttgart, Germany}\\
{\em $^3$Dipartimento di Ingegneria Meccanica e Aerospaziale,\\
Sapienza Universit\`a di Roma, via Eudossiana 18, 00184 Rome, Italy }\\
\end{center}
\vspace{1.2cm}

\renewcommand{\thefootnote}{\arabic{footnote}}
\setcounter{footnote}{0}

\begin{center}
\today
\end{center}

\begin{abstract}
We consider near-critical two-dimensional statistical systems at phase coexistence on the half plane with boundary conditions leading to the formation of a droplet separating coexisting phases. General low-energy properties of two-dimensional field theories are used in order to find exact analytic results for one- and two-point correlation functions of both the energy density and order parameter fields. The subleading finite-size corrections are also computed and interpreted within an exact probabilistic picture in which interfacial fluctuations are characterized by the probability density of a Brownian excursion. The explicit analysis of the closed-form expression for order parameter correlations reveals the long-ranged character of interfacial correlations and their confinement within the interfacial region. The analysis of correlations is then carried out in momentum space through the notion of interface structure factor, which we extend to the case of systems bounded by a flat wall. The presence of the wall and its associated entropic repulsion leads to a specific term in the interface structure factor which we identify.
\end{abstract}

\vfill
$^\natural$squarcio@is.mpg.de, $^\flat$antonio.tinti@uniroma1.it
\end{titlepage}


\section{Introduction}
\label{sec_1}
The study of interfacial phenomena at boundaries is one of the cornerstones of statistical mechanics \cite{gennes_wetting_1985, sullivantelodagama, dietrich_wetting_1988, FLN, BLM, MSchick, BEIMR}. The problem of phase separation and fluctuating interfaces in two dimensions is a particularly interesting one. For long time exact results have been available only for the Ising model \cite{Abraham_review}. This fact is originated by the possibility to find exact diagonalizations of transfer matrices in certain lattice geometries with boundary conditions leading to the formation of interfaces. In this regard, it has to be mentioned the exact solution of the wetting transition with a flat boundary \cite{Abraham1980}, a milestone in the field. These exact results have been crucial for the consolidation of phenomenological interpretations for interfacial behavior in terms of random walks in restricted geometries \cite{fisher_walks_1984} and coarse-grained descriptions treated within path-integral techniques \cite{Burkhardt, VL}.

Despite the wealth of results available for interfaces in the Ising model \cite{Abraham_review}, such an extended degree of knowledge about phase separation for other models has not been achieved until recent times. The principal reason is the obstruction posed by the difficulty of finding exact results for most of the lattice models in certain geometries. On the other hand, it is known that upon approaching a continuous phase transition point the divergence of the bulk correlation length leads to universal behavior and that field theory has proven to be a versatile language for its description in the continuum setting \cite{ZJ,ID}; see \cite{Jasnow_1984, diehl_fieldtheoretical_1986} for boundary field theories in near-critical systems. As a result, the systematic investigation of the tangled scenario of phase separation for arbitrary models must inevitably be formulated within a language that is able to encompass the different bulk universality classes jointly with boundary data; this is the language of field theory.

It has been shown how general results for interface profiles and correlation functions within the interfacial region separating coexisting bulk phases are completely codified by low-energy properties of field theories \cite{DV, DS_twopoint}. The exact theory of phase separation developed in \cite{DV, DS_wetting, DS_wedge, DS_bubbles, DS_wedgebubble, DS_twopoint} has provided a unified framework for the study of interfacial phenomena exhibited by general universality classes in various geometries. Among the various findings, it has been possible to describe boundary wetting transitions \cite{DS_wetting}, wedge filling transitions \cite{DS_wedge}, interfacial wetting \cite{DS_bubbles,DSS1}, and many-body correlation functions \cite{ST_threepoint, ST_fourpoint, Squarcini_Multipoint} for the scaling limit of those models (and boundary conditions) which are not yet solved on the lattice. Moreover, the role of bulk and boundary integrability in these exact findings has been also clarified \cite{DV, DS_bubbles,DS_wedgebubble}.

The concept of interface and the possibility of conferring to it certain fluctuation properties is at the core of effective formulations such as the capillary wave model \cite{BLS_1965, Evans_79}, Weeks' columnar model \cite{Weeks_1977} and subsequent elaborations thereof \cite{BW_1985}. We refer to \cite{PR_NaturePhysics, SREP_2022} for a recent account on the subject. It is known since long time from theory of inhomogeneous fluids that density fluctuations within the interfacial region separating coexisting phases exhibit long-range correlations in the direction parallel to the interface \cite{Wertheim_1976}. This situation happens to be in sharp contrast with the exponential decay of correlations exhibited within pure phases. Going beyond effective formulations, it has been shown \cite{DV, DS_wetting, DS_wedge, DS_bubbles, DS_wedgebubble, DS_twopoint} how a fundamental description of phase separation and interfacial phenomena has to be inevitably formulated in terms of the degrees of freedom of the bulk field theory, which are the asymptotic particle states, as illustrated in this paper.


In this paper, we examine the exact form of correlation functions of both the order parameter and energy density fields for a system bounded on the half-plane. Suitable boundary conditions along the wall are used in order to enforce phase separation in the half-plane through a droplet with pinned endpoints on the wall. Contrary to effective modelings, the approach presented in this paper does not rely on the introduction of the notion of interface but rather follows from the fundamental degrees of freedom of the underlying field theory corresponding to the scaling limit of the statistical system with the appropriate boundary conditions.

More technically, our approach follows as an amalgamation of the field-theoretic formalism developed in \cite{DS_wetting} for the calculation of interface profiles on the half-plane, with the techniques developed in \cite{DS_twopoint} for the calculation of two-point correlation functions. One of the key results of this paper is the analytic expression for the order parameter correlation function in the presence of the fluctuating droplet. The leading asymptotic behavior of the order parameter correlation function in the direction parallel to the interface reads
\begin{equation}
\begin{aligned}
\label{12022021_1651}
\langle \sigma(x,y) \sigma(x,-y) \rangle_{B_{bab}} & = \left( \frac{ \langle \sigma \rangle_{a} + \langle \sigma \rangle_{b} }{ 2 } \right)^{2} + \frac{ \langle \sigma \rangle_{a}^{2} - \langle \sigma \rangle_{b}^{2} }{ 2 } \biggl[ -1 - 4 x \sqrt{\frac{2m}{\pi R}} \textrm{e}^{-\frac{2m}{R}x^{2}} + 2\textrm{erf}\left(\sqrt{\frac{2m}{R}}x\right) \biggr] \\
& + \left( \frac{ \langle \sigma \rangle_{a} - \langle \sigma \rangle_{b} }{ 2 } \right)^{2} \biggl[ 1 - \frac{32mx^{2}}{\pi R} \sqrt{\frac{2y}{R}} \, \textrm{e}^{-\frac{2m}{R}x^{2}} \biggr] + \Os((y/R)^{3/2})
\end{aligned}
\end{equation}
for separations $\xi_{\rm b} \ll y \ll R/2$. In the above, $R$ is the distance between the interface endpoints, $m=1/(2\xi_{\rm b})$, and $\xi_{\rm b}$ is the bulk correlation length. The notation $\langle \cdots \rangle_{B_{bab}}$ stands for statistical averages in the half plane ($x>0$) with boundary conditions enforcing a droplet, and $\langle \cdots \rangle_{a/b}$ stands for expectation values in pure phase $a$ or $b$.

The interface separating phases $a$ and $b$ is characterized by midpoint fluctuations of order $R^{1/2}$ along the $x$-axis. For $R \rightarrow \infty$ the result (\ref{12022021_1651}) reduces to
\begin{equation}
\begin{aligned}
\label{12022021_1652}
\lim_{R \rightarrow \infty} \langle \sigma(x,y) \sigma(x,-y) \rangle_{B_{bab}} & = \langle \sigma \rangle_{a}^{2} \, ,
\end{aligned}
\end{equation}
meaning that unbounded interfacial fluctuations yield an averaging over the phase $a$ enclosed by the droplet. On the other hand, for finite $R$ the most significant variations are localized in the region where it is most probable to find the interface. The term proportional to $\sqrt{y/R}$ is the characteristic signature of the long-range character of density correlations within the interfacial region. Moreover, the $x$-dependent part is amenable of interpretations. The presence of the quadratic factor $x^{2}$, which is also responsible of the entropic repulsion, penalizes correlations in the proximity of the boundary while the Gaussian factor suppresses correlations far away from the boundary. The fact that interfacial fluctuations are long ranged and confined to the interfacial region -- as pointed out by Wertheim \cite{Wertheim_1976} -- is neatly realized by inspection of the analytic result (\ref{12022021_1651}).

Among the various analytic results in real space, we also investigate the long-range character of interfacial correlations in momentum space by extending the notion of interface structure factor to bounded systems. This analysis will allow us to identify in the interface structure factor a specific correction stemming from the entropic repulsion of the interface from the wall.

This paper is structured as follows. As a warmup, in Sec.~\ref{section_2} the calculation of the order parameter profile for a droplet is reviewed. In Sec.~\ref{section_3} the field-theoretical method for the calculation of pair correlation functions is presented in its simplest form, namely the energy density correlations. The calculation of spin-spin correlations is then addressed in Sec.~\ref{section_4} and the connection with a probabilistic interpretation is also illustrated. The interface structure factor for a droplet is defined and calculated in Sec.~\ref{section_6}. Concluding remarks are collected in Sec.~\ref{sec_conclusions}. A series of appendices collects some mathematical details involved in the calculations reported in the main body of the paper.

\section{Magnetization profile}
\label{section_2}
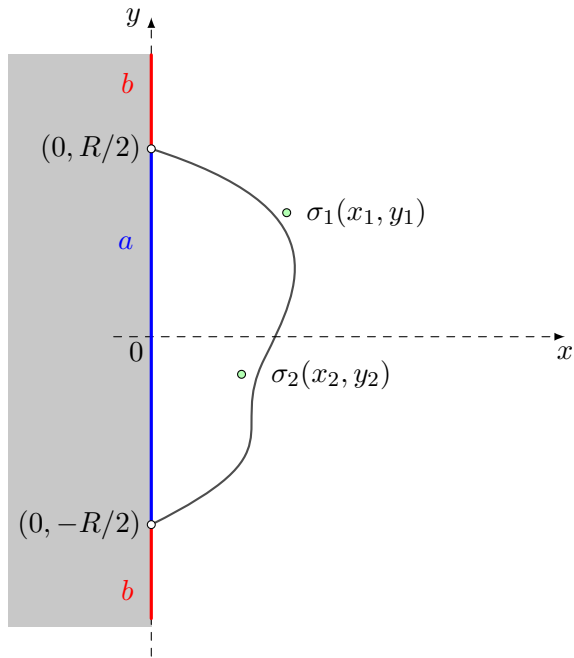
\begin{figure}[htbp]
\centering
\begin{tikzpicture}[thick, line cap=round, >=latex, scale=0.5]
\tikzset{fontscale/.style = {font=\relsize{#1}}}
\shade [left color=mygray, right color=mygray]
(-3.8,-7.7) rectangle (0, 7.5);
\draw[black!70]
(0, -5) ..controls +(4.0, 2.0) and ( $(3, -0.6) + (-1, -2)$ )..
(3, -0.6) ..controls +(1.0, 2) and ( $(0, 5) + (5.0, -1.6)$ )..
(0, 5);
\draw[thin, dashed, ->] (-1, 0) -- (11, 0) node[below] {$x$};
\draw[thin, dashed, ->] (0, -8.5) -- (0, 8.5) node[left] {$y$};
\draw[very thick, red, -] (0, 5) -- (0,7.5) node[] {};
\draw[very thick, red, -] (0, -5) -- (0,-7.5) node[] {};
\draw[very thick, blue, -] (0, -5) -- (0,5) node[] {};
\draw[thin, fill=white] (-0.4, -0.4) circle (0pt) node[] {${\color{black}{0}}$};;
\draw[thin, fill=white] (-0.2, 2.5) circle (0pt) node[left] {${\color{blue}{a}}$};;
\draw[thin, fill=white] (-0.2, 6.75) circle (0pt) node[left] {${\color{red}{b}}$};;
\draw[thin, fill=white] (-0.2, -6.75) circle (0pt) node[left] {${\color{red}{b}}$};;
\draw[thin, fill=white] (0, -5) circle (3pt) node[left] {$(0,-R/2)$};;
\draw[thin, fill=white] (0, 5) circle (3pt) node[left] {$(0,R/2)$};;
\draw[thin, fill=green!30] (3.6, 3.3) circle (3pt) node[right] {\,\,${\sigma_{1}(x_{1},y_{1})}$};;
\draw[thin, fill=green!30] (2.4, -1.0) circle (3pt) node[right] {\,\, ${\sigma_{2}(x_{2},y_{2})}$};;
\end{tikzpicture}
\caption{The half-plane geometry with boundary conditions $B_{bab}$ leading to the formation of a droplet in the halfplane $x>0$ with pinning points in $(0,\pm R/2)$. Green circles indicate the order parameter fields $\sigma_{1}$ and $\sigma_{2}$ appearing in correlation functions considered in this paper.}
\label{fig_geometry}
\end{figure}

We begin by reviewing the calculation of the magnetization profile for a statistical system at a first order phase transition with boundary conditions enforcing a droplet as the one shown in Fig.~\ref{fig_geometry}. The derivation follows closely the original calculation given in \cite{DS_wetting}, which in the present exposition is extended in order to include finite-size corrections originated by interface structure. We consider a two-dimensional system on the first order phase transition line close to a second order phase transition point. The scaling limit in the near-critical region can be described in terms of a Euclidean field theory obtained by analytic continuation to imaginary time of a relativistic quantum field theory in a $1+1$ dimensional space-time. Elementary excitations in $1+1$ dimensional quantum field theories are topological particles (kinks) which interpolate between different ground states and whose propagation in space-time corresponds to boundaries between different coexisting phases.

We study phase separation on the half-plane geometry with boundary conditions along the $y$-axis leading to the formation of a droplet, as illustrated in Fig.~\ref{fig_geometry}. The system is considered for temperatures $T$ such that the bulk correlation length $\xi_{\rm b}$ is much smaller than the separation $R$ between interface endpoints; hence, $\xi_{\rm b} \ll R$. Moreover, we also assume that $\xi_{\rm b}$ is much larger than any microscopic scale\footnote{$a_{0}$ is of the order of the lattice spacing in a Monte Carlo simulation.}, i.e., $\xi_{\rm b} \gg a_{0}$. This last assumption allows us to adopt a field-theoretical language for the description of the system in the regime $a_{0} \ll \xi_{\rm b} \ll R$.

The partition function of the system with $B_{bab}$ boundary conditions can be written as follows
\begin{equation}
\label{e001}
\mathcal{Z} = {}_{B_{b}}\langle 0 \vert \mu_{ba}(0,R/2) \mu_{ab}(0,-R/2) \vert 0 \rangle_{B_{b}} \, ,
\end{equation}
where $\vert 0 \rangle_{B_{b}}$ is the vacuum state in which the boundary has fixed boundary condition with spin in state $b$ and $\mu_{ab}(0, \pm R/2)$ is the boundary condition changing operator which implements the switch of boundary condition from $b$ to $a$ in the point $(0,\pm R/2)$. The matrix element of $\mu_{ba}$ between the vacuum and the single-kink state takes the form
\begin{equation}
\label{e002}
{}_{B_{b}}\langle 0 \vert \mu_{ba}(0,y) \vert K_{ab}(\theta) \rangle = \textrm{e}^{-my} \mathcal{F}_{\mu}(\theta) \, ,
\end{equation}
where $m$ is the kink mass and $\mathcal{F}_{\mu}(\theta)$ is the form factor of the boundary condition changing operator $\mu_{ba}$. Form factors of boundary operators have been studied extensively in the framework of massive integrable quantum field theories \cite{LS_1998, LS, BPT_BFF, BH_2016}. The non-vanishing of the matrix element of $\mu_{ba}$ with the one-kink state corresponds to the pinning of a single\footnote{The pinning of a double kink corresponds to the formation of an intermediate phase \cite{DS_bubbles, DS_wedgebubble}.} domain wall in those points in which the boundary condition switches from $a$ to $b$ \cite{DS_wetting}. By inserting a resolution of the identity between the operators $\mu_{ba}$ and $\mu_{ab}$ in (\ref{e001}) and using the normalization of states, it follows that
\begin{equation}
\label{e003}
\mathcal{Z} = \int_{0}^{\infty} \frac{\textrm{d}\theta}{2\pi} \, |\mathcal{F}_{\mu}(\theta)|^{2} \textrm{e}^{-mR\cosh\theta} + \Os(\textrm{e}^{-2mR})\, .
\end{equation}
The restriction of the integration over positive rapidities is due to the wall. The single-kink state appearing in the first term in the right hand side of (\ref{e003}) gives the dominant contribution to the partition function in the regime $mR\gg1$, which is the one we are interested in. In the low-temperature phase, $m=1/(2\xi_{\rm b}) $, with $\xi_{\rm b}$ the bulk correlation length. The aforementioned relationship is a form of Widom's scaling relation that for the Ising model is exact for all subcritical temperatures which follows from duality \cite{FF67,AGM73,Abraham_review}; see also \cite{AMSV, DS2015}. The regime $R\gg\xi_{\rm b}$ projects the integrand in (\ref{e003}) at small rapidities and the boundary form factor expands as follows
\begin{equation}
\label{e004}
\mathcal{F}_{\mu}(\theta) = \im \mathfrak{a} \theta + \mathfrak{b} \theta^{2} + \Os(\theta^{3}) \, ,
\end{equation}
where the model-dependent (real) coefficients $\mathfrak{a}$ and $\mathfrak{b}$ are known for boundary integrable field theories \cite{LS, BH_2016}. It has been already noticed how the linear behavior at low rapidities exhibited by $\mathcal{F}_{\mu}(\theta)$ is ultimately responsible for the entropic repulsion of the interface from the wall \cite{DS_wetting}. By inserting (\ref{e004}) into (\ref{e003}) a saddle-point calculation yields
\begin{equation}
\label{e005}
\mathcal{Z} = \frac{ \mathfrak{a}^{2} \textrm{e}^{-mR} }{ \sqrt{\pi} (2mR)^{3/2} } + \Os(R^{-5/2}) \, .
\end{equation}

The magnetization profile for $|y|<R/2$ is defined by
\begin{equation}
\label{e002}
\langle \sigma(x,y) \rangle_{B_{bab}} = \frac{ 1 }{ \mathcal{Z} } {}_{B_{b}}\langle 0 \vert \mu_{ba}(0,R/2) \sigma(x,y) \mu_{ab}(0,-R/2) \vert 0 \rangle_{B_{b}} \, ,
\end{equation}
the subscript $B_{bab}$ stands for the expectation value with the boundary conditions illustrated in Fig.~\ref{fig_geometry}. For $mx\gg1$ the spin field entering in (\ref{e002}) can be treated as a bulk field, for which one can use translation invariance on the plane
\begin{equation}
\label{01092020_1654}
\sigma(x,y) = \textrm{e}^{\im x P + y H} \sigma(0,0) \textrm{e}^{-\im x P - y H} 
\end{equation}
in order to bring the field to the origin; $H$ and $P$ are the Hamiltonian and momentum operators in field theory. The expectation value (\ref{e002}) becomes
\begin{equation}
\begin{aligned}
\label{e006}
\langle \sigma(x,y) \rangle_{B_{bab}} & = \frac{1}{\mathcal{Z}} \int_{\mathbb{R}^{2}} \frac{\textrm{d}\theta_{1}\textrm{d}\theta_{2}}{(2\pi)^{2}} \mathcal{F}_{\mu}(\theta_{1}) \mathcal{M}_{ab}^{\sigma}(\theta_{1}\vert\theta_{2}) \mathcal{F}_{\mu}^{*}(\theta_{2}) U(\theta_{1},\theta_{2}) \, ,
\end{aligned}
\end{equation}
with $\mathcal{M}_{ab}^{\sigma}(\theta_{1}\vert\theta_{2}) = \langle K_{ab}(\theta_{1}) \vert \sigma(0,0) \vert K_{ba}(\theta_{2}) \rangle$ and
\begin{equation}
\label{e007}
U(\theta_{1},\theta_{2}) = \exp\biggl[ - m\left( \frac{R}{2} - y \right) \cosh\theta_{1} - m\left( \frac{R}{2} + y \right) \cosh\theta_{2} + \im m x (\sinh\theta_{1}-\sinh\theta_{2}) \biggr] \, .
\end{equation}
The connected part of the one-point correlation function (\ref{e006}), denoted $\langle \sigma(x,y) \rangle_{B_{bab}}^{\rm CP}$, is determined by the connected part of the matrix element $\mathcal{M}_{ab}^{\sigma}(\theta_{1}\vert\theta_{2})$, the latter reads
\begin{equation}
\label{ }
\left( \mathcal{M}_{ab}^{\sigma}(\theta_{1}\vert\theta_{2}) \right)^{\rm CP} = F_{aba}^{\sigma}(\theta_{12}+\im \pi) \, ,
\end{equation}
where $F_{aba}^{\sigma}(\theta_{12}) = \langle 0 \vert \sigma(0,0) \vert K_{ab}(\theta_{1}) K_{ba}(\theta_{2}) \rangle$ is the two-particle (bulk) form factor of the operator $\sigma$ \cite{DS_wetting}. By virtue of relativistic invariance the bulk form factor depends on the rapidities $\theta_{1}$ and $\theta_{2}$ through the difference $\theta_{12} \equiv \theta_{1}-\theta_{2}$ \cite{Smirnov}. On the other hand, boundary form factors do not exhibit such a symmetry \cite{LS, BPT_BFF, BH_2016}. The low-rapidity expansion of the bulk form factor reads
\begin{equation}
\label{lowthetaFF}
F_{aba}^{\sigma}(\theta_{12}+\im \pi) = \sum_{k=-1}^{\infty} c_{k} \theta_{12}^{k} \, .
\end{equation}
The term with $k=-1$ is due to the kinematical pole exhibited by the form factor \cite{Smirnov}. The above expansion is then combined with the corresponding low-rapidity behavior of boundary form factors (\ref{e004}). Hence, one writes
\begin{equation}
\begin{aligned}
\label{e008}
\mathcal{F}_{\mu}(\theta_{1}) F_{aba}^{\sigma}(\theta_{12}+\im \pi) \mathcal{F}_{\mu}^{*}(\theta_{2}) & = c_{-1} \mathfrak{a}^{2} \biggl[   \frac{\theta_{1} \theta_{2}}{\theta_{12}} - \im \omega \theta_{1}\theta_{2} + \Os(\theta^{3}) \biggr] \, ,
\end{aligned}
\end{equation}
where $\omega = \mathfrak{b}/\mathfrak{a} + \im c_{0}/c_{-1}$. The notation $\Os(\theta^{3})$ stands for homogeneous terms with total degree $3$ in the rapidity variables (e.g., $\theta_{1}^{2}\theta_{2}$). By rescaling rapidities through $\theta_{i} \rightarrow \sqrt{2/(mR)} \theta_{i}$ and organizing the result in the form of a power series in the small parameter $(mR)^{-1/2}$, we find
\begin{equation}
\label{e009}
\langle \sigma(x,y) \rangle_{B_{bab}}^{\rm CP} = \frac{ 2c_{-1} }{ \pi^{3/2} } \biggl[ \{\{ \frac{ \theta_{1} \theta_{2} }{ \theta_{12} } \}\} - \im \frac{\sqrt{2} \omega }{\sqrt{mR}} \{\{ \theta_{1} \theta_{2} \}\} \biggr] + \Os\left( (mR)^{-1} \right) \, ;
\end{equation}
the notation $\{\{ g(\theta_{1},\theta_{2}) \}\}$ stands for the contribution of $g(\theta_{1},\theta_{2})$ to the magnetization profile, with $g(\theta_{1},\theta_{2})$ the generic term of the expansion (\ref{e008}). The double curly bracket notation is defined as follows:
\begin{equation}
\begin{aligned}
\label{}
\{\{ g(\theta_{1},\theta_{2}) \}\} & = \dashint_{\mathbb{R}^{2}}\textrm{d}\theta_{1}\textrm{d}\theta_{2} \, g(\theta_{1},\theta_{2}) Y(\theta_{1},\theta_{2}) \, , \\
Y(\theta_{1},\theta_{2}) & = \exp\biggl[ - \frac{1-\tau}{2} \theta_{1}^{2} - \frac{1+\tau}{2} \theta_{2}^{2} + \im \eta \theta_{12} \biggr] \, .
\end{aligned}
\end{equation}
The symbol $\dashint$ stands for the principal value of the integral, which is actually needed since spin field matrix elements exhibit a kinematical pole \cite{DV}. Then, $\eta$ and $\tau$ are the rescaled coordinates defined by
\begin{equation}
\label{03092020_1540}
\eta = x/\lambda \, , \qquad \lambda = \sqrt{R/(2m)} \, , \qquad \tau=2y/R \, .
\end{equation}
A simple calculation yields
\begin{equation}
\begin{aligned}
\label{e011}
\{\{ \frac{ \theta_{1} \theta_{2} }{ \theta_{12} } \}\} & = \frac{ \im \pi^{3/2}}{2} \mathcal{D}(\chi) \, , \qquad \{\{ \theta_{1}\theta_{2} \}\} & = \frac{ \pi^{3/2} \lambda }{ 2 } P_{1}(x,y) \, ,
\end{aligned}
\end{equation}
where
\begin{equation}
\begin{aligned}
\label{e012}
\mathcal{D}(\chi) & = - \frac{ 2 }{ \sqrt{\pi} } \chi \textrm{e}^{-\chi^{2} } + \textrm{erf}(\chi) \, , \qquad P_{1}(x,y) & = \frac{4 \chi^{2}}{\sqrt{\pi}\kappa\lambda} \textrm{e}^{-\chi^{2}} \, ,
\end{aligned}
\end{equation}
and
\begin{equation}
\label{03092020_1540}
\kappa = \sqrt{1-\tau^{2}} \, , \qquad \qquad \chi=x/(\kappa\lambda) \, .
\end{equation}
In order to compute $\{\{ \theta_{1}\theta_{2}/\theta_{12} \}\}$ it is convenient to remove the kinematical pole singularity $1/\theta_{12}$ by taking the first derivative with respect to $\eta$ and thus compute $\{\{ \theta_{1} \theta_{2} \}\}$. Then, by integrating back to $\eta$, we find the desired result. The integration constant generated by such a procedure can be set to zero and eventually it can be reabsorbed into the disconnected term of the matrix element. Such term originates the offset for the profile which is uniquely fixed by the asymptotic boundary conditions: $\langle \sigma(x \rightarrow + \infty, y ) \rangle_{B_{bab}} = \langle \sigma \rangle_{b}$.

Thanks to the above results, and using the known expression for the residue of the two-particle form factor of the spin field, $c_{-1} = \im (\langle \sigma \rangle_{a} - \langle \sigma \rangle_{b})$ \cite{DC98,DV}, the magnetization profile (\ref{e009}) becomes
\begin{equation}
\label{e011}
\langle \sigma(x,y) \rangle_{B_{bab}} = \langle \sigma \rangle_{a} - \Bigl[ \langle \sigma \rangle_{a} - \langle \sigma \rangle_{b} \Bigr] \mathcal{D}(\chi) + \omega \Bigl[ \langle \sigma \rangle_{a} - \langle \sigma \rangle_{b} \Bigr] P_{1}(x;y)/m + \Os(R^{-1}) \, , \quad mx \gg1 \, .
\end{equation}
Since the term proportional to $\omega$ is of order $(\xi_{\rm b}/R)^{1/2}$, it is a subleading correction to the magnetization profile. The droplet profile $\mathcal{D}(\chi)$  is a monotonous function which interpolates between $0$ and $+1$ as $x$ varies from $x=0$ to $x \rightarrow + \infty$ with $|y|<R/2$. Since the leading order term in (\ref{e011}) depends on $x$ and $y$ through $\chi$, it follows that constant values of $\chi$ give the contour lines of the magnetization profile in the plane. As a result, these contour lines are arcs of ellipses with implicit equation
\begin{equation}
\label{15022021_0837}
\frac{x^{2}}{cR\xi_{\rm b}} + \frac{4y^{2}}{R^{2}} = 1 \, ,
\end{equation}
corresponding to $\chi=c$. The theoretical prediction (\ref{15022021_0837}), which is known for the Ising model \cite{Abraham1980}, has been found also within SOS models \cite{Burkhardt, VL} and tested again numerical simulations \cite{Selke_1989_56}. The result (\ref{15022021_0837}) implies that the contour line $\chi=c$ crosses the horizontal axis in $x=\sqrt{c R \xi_{\rm b}}$. The proportionality to $\sqrt{R}$ admits a physical interpretation in terms of a one-dimensional random walk on the half-line, as we are going to show in a while.

\begin{figure}[htbp]
\centering
\includegraphics[width=105mm]{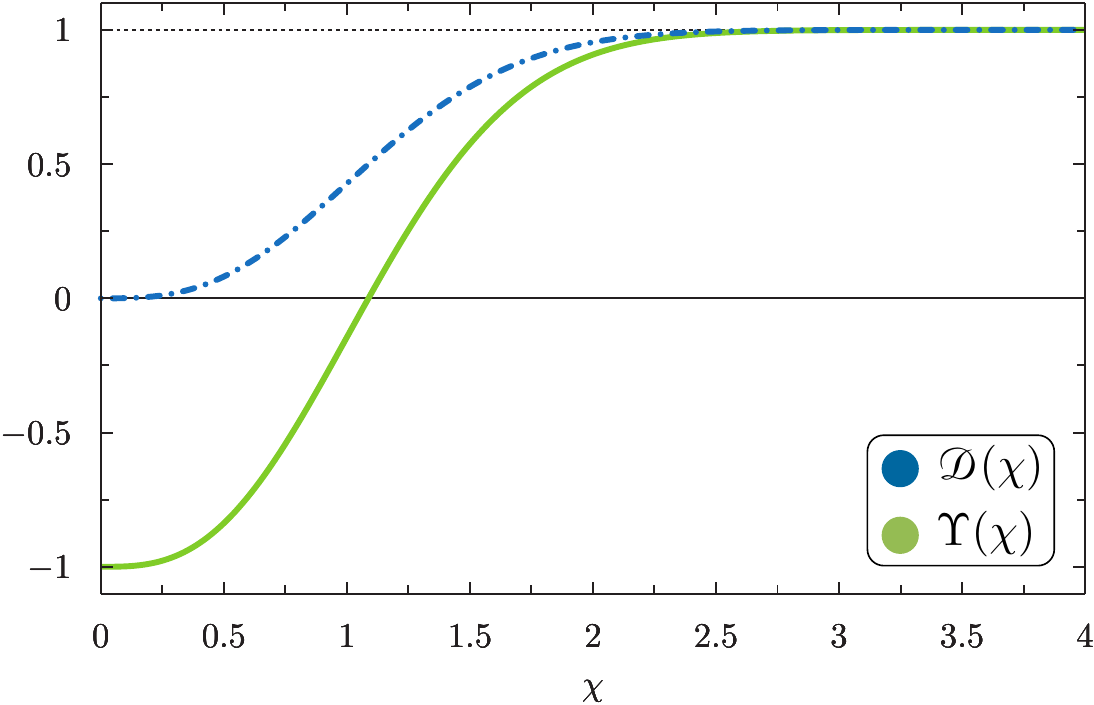}
\caption{The droplet profiles $\mathcal{D}(\chi)$ and $\Upsilon(\chi)$.}
\label{fig_drops}
\end{figure}

For lather convenience, we also introduce the droplet profile $\Upsilon(\chi) = 2\mathcal{D}(\chi)-1$,
\begin{equation}
\label{ }
\Upsilon(\chi) = -1 - \frac{ 4 }{ \sqrt{\pi} } \chi \textrm{e}^{-\chi^{2} } + 2\textrm{erf}(\chi) \, ,
\end{equation}
which interpolates between $-1$ and $+1$; both the profiles are plotted in Fig.~\ref{fig_drops}.

It is instructive to specialize the general result (\ref{e011}) to the explicit case of the Ising model. From the known expression of the boundary form factor \cite{LS}, we extract $\omega_{\rm Ising}=-1/2$ and therefore
\begin{equation}
\label{0933}
\langle \sigma(x,y) \rangle_{B_{+-+}} = M\Upsilon(\chi) + \frac{ 2^{5/2}M }{ \sqrt{\pi mR}\kappa } \chi^{2}\textrm{e}^{-\chi^{2} }+ \Os(R^{-1}) \, ,
\end{equation}
where $M=\langle \sigma \rangle_{+}=-\langle \sigma \rangle_{-}>0$ is the spontaneous magnetization. The leading-order term $\propto \Upsilon(\chi)$ shown in (\ref{0933}) coincides with scaling limit of exact results obtained for the square lattice Ising model \cite{Abraham1980, AI, Abraham_review} and from the path-integral formulation of Solid-On-Solid models \cite{VL,Burkhardt}.

Within a probabilistic interpretation \cite{DS_wetting,DS_wedge} the magnetization profile can be derived by summing over interfacial configurations weighted with a certain passage probability density $P_{1}(x,y)$. Thus, the magnetization profile is computed as follows
\begin{equation}
\label{proba}
\langle \sigma(x,y) \rangle_{B_{bab}} = \int_{0}^{\infty}\textrm{d}u \, P_{1}(u,y) \sigma_{ab}(x \vert u) \, ,
\end{equation}
where
\begin{equation}
\label{sharp}
\sigma_{ab}(x \vert u) = \langle \sigma \rangle_{a} - \left( \langle \sigma \rangle_{a} - \langle \sigma \rangle_{b} \right) \theta(x-u) + A_{ab} \delta(x-u) + \dots \, ,
\end{equation}
gives the magnetization at point $x$ when the interface, regarded as a sharp curve, passes through $(u,u+\textrm{d}u)$ at ordinate $y$. The first two terms in (\ref{sharp}) corresponds to a description in which the coexisting phases are sharply separated. The correction to this picture is achieved by endowing the structureless sharp profile with interface structure effects $\propto A_{ab}$ whose effect is to produce subleading finite-size corrections \cite{DV}. The identification of the passage probability with the expression for $P_{1}(x,y)$ given in (\ref{e012}) is established by matching the probabilistic construction (\ref{proba}) with the field-theoretic result (\ref{e011}). The following identities prove to be useful
\begin{equation}
\begin{aligned}
\label{droplet_profile}
\int_{0}^{x}\textrm{d}u \, P_{1}(u,y) & = \mathcal{D}(\chi) \, , \\
\int_{0}^{\infty}\textrm{d}u \, P_{1}(u,y) \textrm{sign}(x-u) & = \Upsilon(\chi) \, ,
\end{aligned}
\end{equation}
jointly with the fact that $\int_{0}^{\infty}\textrm{d}u \, P_{1}(u,y)=1$, i.e., the passage probability density $P_{1}(x,y)$ is normalized. The above passage probability characterizes the so called \emph{Brownian excursion} and its expression is given by (\ref{e012}). By pushing the comparison at the next-to-leading order, we can identify the structure amplitude
\begin{equation}
\begin{aligned}
\label{01092020_2047}
A_{ab} & = \frac{\omega}{m} \Bigl[ \langle \sigma \rangle_{a} - \langle \sigma \rangle_{b} \Bigr] \\
& = \frac{c_{0}}{m} + \frac{\mathfrak{b}}{\mathfrak{a}} \frac{\Delta\langle\sigma\rangle}{m} \, .
\end{aligned}
\end{equation}
The application of the above reasoning to the profile of the energy density field on the half plane follows \emph{mutatis mutandis} (we refer to \cite{Squarcini_Multipoint, ST_threepoint} for the calculation on the strip geometry). It is worth noticing that on the strip $A_{ab}^{\rm (strip)}=c_{0}/m$ vanishes for the Ising model while the corresponding result for the half-plane geometry given by (\ref{01092020_2047}) is actually non zero. Thus, $A_{ab}$ is geometry-dependent.

\section{Energy density correlations}
\label{section_3}
The connected part of energy density correlation functions contains information on the passage probability. More precisely, the connected energy density correlation function is proportional to the joint passage probability. Such a feature, which has been established for the strip geometry \cite{Squarcini_Multipoint}, is valid also for the half-plane, as we are going to show. Let $P_{2}(x_{1},y_{1};x_{2},y_{2})$ be the joint passage probability density, therefore $P_{2}(x_{1},y_{1};x_{2},y_{2})\textrm{d}x_{1} \textrm{d}x_{2}$ is the probability of the interface to pass through the intervals $(x_{1}, x_{1}+\rd x_{1})$ \emph{and} $(x_{2}, x_{2}+\rd x_{2})$ at ordinates $y_{1}$ and $y_{2}$, respectively.

In this section, we extract $P_{2}(x_{1},y_{1};x_{2},y_{2})$ from the two-point correlation function of the energy density field computed in field theory. The correlation function we are interested in reads
\begin{equation}
\label{31}
\langle \varepsilon(x_{1},y_{1}) \varepsilon(x_{2},y_{2}) \rangle_{ B_{bab} }^{\rm CP} = \frac{ 1 }{ \mathcal{Z} } {}_{B_{b}}\langle 0 \vert \mu_{ba}(0,R/2) \varepsilon(x_{1},y_{1}) \varepsilon(x_{2},y_{2}) \mu_{ab}(0,-R/2) \vert 0 \rangle_{B_{b}}^{\rm CP} \, ;
\end{equation}
with $y_{1}-y_{2} \gg \xi_{\rm b}$ and both energy density fields far from the boundaries in the following sense: $(R/2)-y_{1} \gg \xi_{\rm b}$ and $y_{2} +R/2 \gg \xi_{\rm b}$. The superscript CP means that we retain the connected part. By employing the decomposition over intermediate states, (\ref{31}) becomes
\begin{equation}
\begin{aligned}
\label{32}
\langle \varepsilon(x_{1},y_{1}) \varepsilon(x_{2},y_{2}) \rangle_{ B_{bab} }^{\rm CP} & = \frac{1}{\mathcal{Z}} \int \frac{\textrm{d}\theta_{1}\textrm{d}\theta_{2}\textrm{d}\theta_{3}}{(2\pi)^{3}} \biggl[ \mathcal{F}_{\mu}(\theta_{1}) \left( \mathcal{M}_{ab}^{\varepsilon}(\theta_{1}\vert \theta_{2}) \mathcal{M}_{ab}^{\varepsilon}(\theta_{2} \vert \theta_{3}) \right)^{\rm CP} \mathcal{F}_{\mu}^{*}(\theta_{3}) U_{+}(\{\theta\}) \\
& + \mathcal{F}_{\mu}(\theta_{1}) \left( \mathcal{M}_{ab}^{\varepsilon}(\theta_{1} \vert -\theta_{2}) \mathcal{R}_{aba}(\theta_{2}) \mathcal{M}_{ab}^{\varepsilon}(\theta_{2} \vert \theta_{3}) \right)^{\rm CP} \mathcal{F}_{\mu}^{*}(\theta_{3}) U_{-}(\{\theta\}) \biggr] \, ,
\end{aligned}
\end{equation}
where
\begin{equation}
\begin{aligned}
\label{03092020_1358}
U_{\pm}(\{\theta\}) & = \exp\biggl[ -m\left( \frac{R}{2} - y_{1} \right) \cosh\theta_{1}-m (y_{1}-y_{2}) \cosh\theta_{2}-m \left( y_{2} + \frac{R}{2} \right) \cosh\theta_{3} \\
& + \im m x_{1} ( \sinh\theta_{1} \mp \sinh\theta_{2} ) + \im m x_{2} (\sinh\theta_{2} - \sinh\theta_{3} ) \biggr] \, .
\end{aligned}
\end{equation}
The second term appearing in the right hand side of (\ref{32}) contains the boundary $S$-matrix\footnote{See \cite{GZ,Chim_1995,Chim_1996} for boundary $S$-matrices in integrable field theories.} $\mathcal{R}_{aba}(\theta_{2})$ which gives the amplitude for the scattering of a kink state off the vertical wall with fixed boundary conditions $a$ \cite{GZ}; the latter admits the pictorial representation of (\ref{e034}).
\begin{equation}
\label{e034}
\mathcal{R}_{aba}(\theta)
=
\begin{tikzpicture}[baseline={([yshift=-.6ex]current bounding box.center)},vertex/.style={anchor=base, circle, minimum size=50mm, inner sep=0pt}]
\draw[-,black]   (-1.8, 0) -- (0, 1.8);
\draw[->,black]   (-1.8, 0) -- (-0.9, 0.9);
\draw[-,black]   (-1.8, 0) -- (0, -1.8);
\draw[->,black]   (0, -1.8) -- (-0.9, -0.9);
\path (0,0.0) node () {} (-1.4, 1.2) node (a) { ${\color{blue}{a}}$ };
\path (0,0.0) node () {} (-1.4, -1.2) node (b) { ${\color{blue}{a}}$ };
\path (0,0.0) node () {} (-1.4, 0) node (b) { ${\color{red}{b}}$ };
\path (0,0.0) node () {} (-0.15+0.1, -1.3) node (b) { ${\color{black}{\theta}}$ };
\path (0,0.0) node () {} (-0.30+0.1, 1.3) node (b) { ${\color{black}{-\theta}}$ };
\shade [left color=white, right color=gray]
(-2.0,-1.8) rectangle (-1.8, 1.8);
\end{tikzpicture}
\end{equation}
The two terms appearing in the square brackets of (\ref{32}) admit the diagrammatic representation provided in (\ref{23012021_1730}).
\begin{equation}
\begin{aligned}
\label{23012021_1730}
\mathfrak{M}_{D}^{\varepsilon, {\rm CP}} = 
\begin{tikzpicture}[baseline={([yshift=-.6ex]current bounding box.center)},vertex/.style={anchor=base, circle, minimum size=50mm, inner sep=0pt}]
\path (0.2, 0.8) node[circle, draw, fill=myyellow!60] (s1) {$\varepsilon$} (-0.2, -0.8) node[circle, draw, fill=myyellow!60](s2) {$\varepsilon$};
\draw[-,black]   (s1) -- (s2);
\draw[-,black]   (s1) -- (0.2, 2);
\draw[-,black]   (s2) -- (-0.2, -2);
\path (0,0.0) node () {} (-0.75,0) node (s4) { ${\color{blue}{a}}$ };
\path (0,0.0) node () {} (0.75,0) node (s4) { ${\color{red}{b}}$ };
\path (0,0.0) node () {} (0.45,1.6) node (s4) { ${\color{black}{\theta_{1}}}$ };
\path (0,0.0) node () {} (0.25,0.0) node (s4) { ${\color{black}{\theta_{2}}}$ };
\path (0,0.0) node () {} (0.45-0.4,-1.6) node (s4) { ${\color{black}{\theta_{3}}}$ };
\shade [left color=white, right color=gray]
(-2.2+0.6,-2.0) rectangle (-2.0+0.6, 2.0);
\end{tikzpicture}
\quad
& = F_{aba}^{\varepsilon}(\theta_{12}+\im \pi) F_{aba}^{\varepsilon}(\theta_{23}+\im \pi) \\
\mathfrak{M}_{R}^{\varepsilon, {\rm CP}} = 
\begin{tikzpicture}[baseline={([yshift=-.6ex]current bounding box.center)},vertex/.style={anchor=base, circle, minimum size=50mm, inner sep=0pt}]
\path (0.2, 0.8) node[circle, draw, fill=myyellow!60] (s1) {$\varepsilon$} (-0.2, -0.8) node[circle, draw, fill=myyellow!60](s2) {$\varepsilon$};
\draw[-,black]   (s1) -- (0.2, 2);
\draw[-,black]   (s2) -- (-0.2, -2);
\path (0,0.0) node () {} (-0.85, 1.0) node (s4) { ${\color{blue}{a}}$ };
\path (0,0.0) node () {} (-0.85, -1.0) node (s4) { ${\color{blue}{a}}$ };
\path (0,0.0) node () {} (0.75,0) node (s4) { ${\color{red}{b}}$ };
\path (0,0.0) node () {} (0.45,1.6) node (s4) { ${\color{black}{\theta_{1}}}$ };
\path (0,0.0) node () {} (-0.71, -0.45) node (s4) { ${\color{black}{\theta_{2}}}$ };
\path (0,0.0) node () {} (-0.86, 0.45) node (s4) { ${\color{black}{-\theta_{2}}}$ };
\path (0,0.0) node () {} (0.45-0.4,-1.6) node (s4) { ${\color{black}{\theta_{3}}}$ };

\draw[-,black]   (0.2, 0.5) -- (0.2, 0.4);
\draw[-,black]   (0.2, 0.4) -- (-1.4, 0);
\draw[-,black]   (-0.2, -0.4) -- (-1.4, 0);
\draw[-,black]   (-0.2, -0.5) -- (-0.2, -0.4);


\shade [left color=white, right color=gray]
(-2.2+0.6,-2.0) rectangle (-2.0+0.6, 2.0);
\end{tikzpicture}
\quad
& = F_{aba}^{\varepsilon}(\widehat{\theta}_{12}+\im \pi) \mathcal{R}_{aba}(\theta_{2})F_{aba}^{\varepsilon}(\theta_{23}+\im \pi) \, , \qquad \widehat{\theta}_{ij} = \theta_{i} + \theta_{j} \, .
\end{aligned}
\end{equation}
The kink with rapidity $\theta_{2}$ appearing in the diagram $\mathfrak{M}_{D}^{\varepsilon, {\rm CP}}$ connects the two energy density fields in a ``direct'' ($D$) fashion while in $\mathfrak{M}_{R}^{\varepsilon, {\rm CP}}$ the kink with rapidity $\theta_{2}$ is reflected ($R$) off the boundary before being connected to the other field. As a result, $\mathfrak{M}_{R}^{\varepsilon, {\rm CP}}$ picks up a boundary reflection factor $\mathcal{R}_{aba}$. The leading-order form of the correlation function (\ref{31}) emerges from the behavior of matrix elements at small momenta. Since $\mathcal{R}_{aba}(\theta_{2}) = -1 + \Os(\theta_{2})$ for $\theta_{2} \rightarrow 0$, the matrix elements in (\ref{23012021_1730}) tend to $\pm (F_{aba}^{\varepsilon}(\im \pi))^{2}$, respectively. The calculation involved in (\ref{31}) requires the evaluation of Gaussian integrations. Leaving the details in Appendix \ref{appA} and \ref{appB}, we find
\begin{equation}
\begin{aligned}
\label{37}
\langle \varepsilon(x_{1},y_{1}) \varepsilon(x_{2},y_{2}) \rangle_{ B_{bab} }^{\rm CP} & = \frac{ \left( F_{aba}^{\varepsilon}(\im \pi) \right)^{2}  }{ m^{2} } P_{2}(x_{1},y_{1}; x_{2},y_{2}) \, ,
\end{aligned}
\end{equation}
where
\begin{equation}
\label{38}
P_{2}(x_{1},y_{1};x_{2},y_{2}) = \frac{ 8 \chi_{1} \chi_{2} }{ \rho \kappa_{1} \kappa_{2} \lambda^{2} } \bigl[ \Pi_{2}(\sqrt{2}\chi_{1},\sqrt{2}\chi_{2} \vert \rho) - \Pi_{2}(\sqrt{2}\chi_{1},\sqrt{2}\chi_{2} \vert -\rho) \bigr] \, ;
\end{equation}
in the above, $\Pi_{2}(u_{1},u_{2} \vert \rho)$ is a normal Gaussian bivariate distribution (see Appendix \ref{appB}) with correlation coefficient
\begin{equation}
\label{03092020_1250}
\rho = \sqrt{ \frac{ 1 - \tau_{1} }{ 1 + \tau_{1} } \frac{ 1 + \tau_{2} }{ 1 - \tau_{2} } } \, , \qquad \tau_{j}=2y_{j}/R \, ,
\end{equation}
and the notation
\begin{equation}
\label{ }
\chi_{j} = \eta_{j}/\kappa_{j} \, , \qquad \eta_{j}=x_{j}/\lambda \, , \qquad \kappa_{j}=\sqrt{1-\tau_{j}^{2}} 
\end{equation}
has been adopted. Notice that $0 < \rho < 1$, however, since the two fields are both far from each other and far from the boundaries the extremal values ($\rho=0$ and $\rho=1$) are never reached. 

In order to check that (\ref{38}) is indeed the joint passage probability, we primarily observe that (\ref{38}) satisfies the following properties:
\begin{equation}
\begin{aligned}
\label{}
\int_{0}^{\infty}\textrm{d}x_{2} \, P_{2}(x_{1},y_{1};x_{2},y_{2}) & = P_{1}(x_{1},y_{1}) \, , \\
\int_{0}^{\infty}\textrm{d}x_{1} \int_{0}^{\infty}\textrm{d}x_{2} \, P_{2}(x_{1},y_{1};x_{2},y_{2}) & = 1 \, ,
\end{aligned}
\end{equation}
thus, $P_{2}$ is correctly normalized and its marginal reduces to the passage probability $P_{1}$, as consistency requires for passage probabilities.
\begin{figure}[htbp]
\centering
\includegraphics[width=95mm]{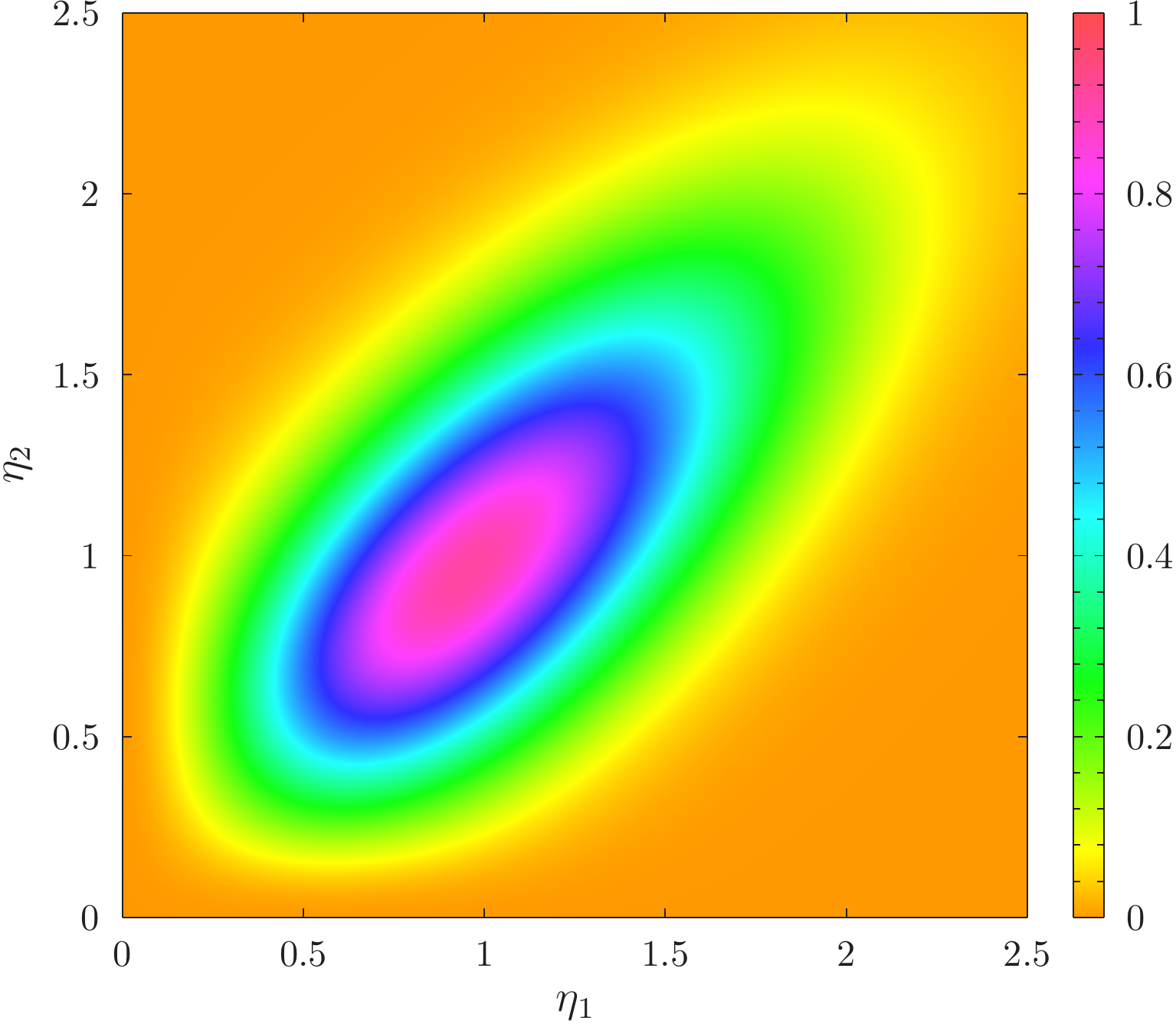}
\caption{The rescaled passage probability $\lambda^{2}P_{2}(x_{1},y;x_{2},-y)$ as function of the rescaled coordinates $\eta_{1}=x_{1}/\lambda$, $\eta_{2}=x_{2}/\lambda$ for $2y/R=0.1$.}
\label{fig_pp}
\end{figure}
Moreover, by applying the above properties the result (\ref{37}) for the energy density correlation function follows from the probabilistic reconstruction
\begin{equation}
\label{310}
\langle \varepsilon(x_{1},y_{1}) \varepsilon(x_{2},y_{2}) \rangle_{ B_{bab} } = \int_{0}^{\infty}\textrm{d}u_{1} \int_{0}^{\infty}\textrm{d}u_{2} \, P_{2}(u_{1},y_{1};u_{2},y_{2}) \varepsilon( x_{1} \vert u_{1} ) \varepsilon( x_{2} \vert u_{2} ) \, ,
\end{equation}
with the energy density profile $\varepsilon( x_{i} \vert u_{i} ) = \langle \varepsilon \rangle + A_{\varepsilon} \delta( x_{i} - u_{i} )+ \dots $ constructed by following the same guidelines which lead us to the magnetization profile. By matching the connected part of (\ref{310}) with the field-theoretical result (\ref{37}), we identify the structure coefficient $A_{\varepsilon} = F_{aba}^{\varepsilon}(\im \pi)/m$. The latter perfectly coincides with the corresponding result already obtained for $n$-point point correlation functions on the strip geometry\footnote{We refer to \cite{ST_threepoint,ST_fourpoint} for analytical and numerical results for $n$-point correlation functions with $n=1,2,3$, and $4$.} \cite{Squarcini_Multipoint}.

\section{Pair correlation function of the spin field}
\label{section_4}
By following the method illustrated in the previous section, we compute the pair correlation function of the spin field. The latter is defined by
\begin{equation}
\label{02092020_1054}
\langle \sigma_{1}(x_{1},y_{1}) \sigma_{2}(x_{2},y_{2}) \rangle_{ B_{bab} } = \frac{ 1 }{ \mathcal{Z} } {}_{B_{b}}\langle 0 \vert \mu_{ba}(0,R/2) \sigma_{1}(x_{1},y_{1}) \sigma_{2}(x_{2},y_{2}) \mu_{ab}(0,-R/2) \vert 0 \rangle_{B_{b}} \, .
\end{equation}
We can replace $\varepsilon$ with $\sigma_{j}$ into (\ref{32}) and use an analogous decomposition for spin fields matrix elements
\begin{equation}
\begin{aligned}
\label{01072020_01}
\mathcal{M}^{\sigma_{1}}(\theta_{1} \vert \theta_{2}) \mathcal{M}^{\sigma_{2}}(\theta_{2} \vert \theta_{3})
&
=
\underbrace{
\begin{tikzpicture}[baseline={([yshift=-.6ex]current bounding box.center)},vertex/.style={anchor=base, circle, minimum size=50mm, inner sep=0pt}]
\path (0.2, 0.9) node[circle, draw, fill=green!30] (s1) {$\sigma_{1}$} (-0.2, -0.9) node[circle, draw, fill=green!30](s2) {$\sigma_{2}$};
\draw[-,black]   (s1) -- (s2);
\draw[-,black]   (s1) -- (0.2, 2);
\draw[-,black]   (s2) -- (-0.2, -2);
\path (0,0.0) node () {} (-0.75,0) node (s4) { ${\color{blue}{a}}$ };
\path (0,0.0) node () {} (0.75,0) node (s4) { ${\color{red}{b}}$ };
\path (0,0.0) node () {} (0.45,1.6) node (s4) { ${\color{black}{\theta_{1}}}$ };
\path (0,0.0) node () {} (0.25,0.0) node (s4) { ${\color{black}{\theta_{2}}}$ };
\path (0,0.0) node () {} (0.45-0.4,-1.6) node (s4) { ${\color{black}{\theta_{3}}}$ };
\shade [left color=white, right color=gray]
(-2.2+0.6,-2.0) rectangle (-2.0+0.6, 2.0);
\end{tikzpicture}
}
_{\mathfrak{M}_{D}^{\rm CP}}
\quad
+
\quad
\underbrace{
\begin{tikzpicture}[baseline={([yshift=-.6ex]current bounding box.center)},vertex/.style={anchor=base, circle, minimum size=50mm, inner sep=0pt}]
\path (0.2, 0.9) node[circle, draw, fill=green!30] (s1) {$\sigma_{1}$} (-0.2, -0.9) node[circle, draw, fill=green!30](s2) {$\sigma_{2}$};
\draw[-,black]   (s1) -- (0.2, 2);
\draw[-,black]   (s2) -- (-0.2, -2);
\path (0,0.0) node () {} (-0.85, 1.0) node (s4) { ${\color{blue}{a}}$ };
\path (0,0.0) node () {} (-0.85, -1.0) node (s4) { ${\color{blue}{a}}$ };
\path (0,0.0) node () {} (0.75,0) node (s4) { ${\color{red}{b}}$ };
\path (0,0.0) node () {} (0.45,1.6) node (s4) { ${\color{black}{\theta_{1}}}$ };
\path (0,0.0) node () {} (-0.71, -0.45) node (s4) { ${\color{black}{\theta_{2}}}$ };
\path (0,0.0) node () {} (-0.86, 0.45) node (s4) { ${\color{black}{-\theta_{2}}}$ };
\path (0,0.0) node () {} (0.45-0.4,-1.6) node (s4) { ${\color{black}{\theta_{3}}}$ };
\draw[-,black]   (0.2, 0.5) -- (0.2, 0.4);
\draw[-,black]   (0.2, 0.4) -- (-1.4, 0);
\draw[-,black]   (-0.2, -0.4) -- (-1.4, 0);
\draw[-,black]   (-0.2, -0.5) -- (-0.2, -0.4);
\shade [left color=white, right color=gray]
(-2.2+0.6,-2.0) rectangle (-2.0+0.6, 2.0);
\end{tikzpicture}
}_{\mathfrak{M}_{R}^{\rm CP}}
\quad
+
\quad
\textrm{disconnected} \, .
\end{aligned}
\end{equation}
The leading low-energy behavior of the above diagrams is given by
\begin{eqnarray}
\mathfrak{M}_{D}^{\sigma, {\rm CP}} & = & (\im)^{2} \Delta\langle \sigma_{1} \rangle\Delta\langle \sigma_{2} \rangle \frac{ 1 }{ \theta_{12}\theta_{23} } \, , \\
\mathfrak{M}_{R}^{\sigma, {\rm CP}} & = & (\im)^{2} \Delta\langle \sigma_{1} \rangle\Delta\langle \sigma_{2} \rangle \frac{ \mathcal{R}_{aba}(0) }{ \widehat{\theta}_{12} \theta_{23} } \, ,
\end{eqnarray}
respectively for the first and second diagrams in (\ref{01072020_01}). In the above, $\mathcal{R}_{aba}(0)$ is the boundary $S$-matrix evaluated at zero rapidity. It has to be noticed that within the low-rapidity regime pertinent to $mR\gg1$ the boundary $S$-matrix reduces to $\mathcal{R}_{aba}(0)=-1$. Such a limiting behavior suffices for the determination of correlation functions at leading order in powers of $(\xi_{\rm b}/R)^{1/2}$.

The calculation of (\ref{02092020_1054}) proceeds as follows. Let us indicate with $\{ \mathfrak{M} \}$ the contribution of the diagram $\mathfrak{M}$ to the pair correlation function. The two diagrams appearing in (\ref{01072020_01}) give the following contribution to the spin-spin correlation function
\begin{equation}
\label{ }
\{ \mathfrak{M}_{D}^{\sigma, {\rm CP}} + \mathfrak{M}_{R}^{\sigma, {\rm CP}} \} = - \frac{ \Delta\langle\sigma_{1}\rangle  \Delta\langle\sigma_{2}\rangle }{ \pi^{5/2} } \biggl[ \Lbag \frac{\theta_{1}\theta_{3}}{\theta_{12}\theta_{23}} \Rbag_{+} - \Lbag \frac{\theta_{1}\theta_{3}}{\widehat{\theta}_{12}\theta_{23}} \Rbag_{-} \biggr] \, ,
\end{equation}
where $\Lbag \cdots \Rbag_{\pm}$ is the notation defined in (\ref{02092020_1057}). By taking the first derivatives with respect to $x_{1}$ and $x_{2}$ and using the identity (\ref{A001}), which relates the calculation of matrix elements to the joint passage probability, it follows that
\begin{equation}
\begin{aligned}
\label{}
\partial_{ x_{1} } \partial_{ x_{2} } \{ \mathfrak{M}_{D}^{\sigma, {\rm CP}} + \mathfrak{M}_{R}^{\sigma, \rm CP} \} & = \frac{ \Delta\langle\sigma_{1}\rangle  \Delta\langle\sigma_{2}\rangle }{ \pi^{5/2} \lambda^{2}} \biggl[ \Lbag \theta_{1}\theta_{3} \Rbag_{+} - \Lbag \theta_{1}\theta_{3} \Rbag_{-} \biggr] \, , \\
& = \Delta\langle\sigma_{1}\rangle  \Delta\langle\sigma_{2}\rangle P_{2}(x_{1},y_{1};x_{2},y_{2}) \, .
\end{aligned}
\end{equation}
Thus, by integrating back with respect to $x_{1}$ and $x_{2}$, we have
\begin{equation}
\label{ }
\{ \mathfrak{M}_{D}^{\sigma, {\rm CP}} + \mathfrak{M}_{R}^{\sigma, {\rm CP}} \} = \Delta\langle\sigma_{1}\rangle \Delta\langle\sigma_{2}\rangle \int_{0}^{x_{1}}\textrm{d}u_{1} \int_{0}^{x_{2}}\textrm{d}u_{2} \, P_{2}(u_{1},y_{1};u_{2},y_{2}) + \Os(R^{-1/2}) \, ,
\end{equation}
up to integration terms\footnote{The integration constants are actually functions which are annihilated by the differential operator $\partial_{x_{1}}\partial_{x_{2}}.$} which, without loss of generality, can be reabsorbed in the disconnected diagrams; for example, by fixing their contribution for $x_{j} \rightarrow +\infty$.

The disconnected terms appearing in the right hand side of (\ref{01072020_01}) are depicted as follows:
\begin{equation}
\begin{aligned}
\label{e048}
\mathfrak{M}^{\sigma_{1}, {\rm disc}} & = \quad 
\begin{tikzpicture}[baseline={([yshift=-.6ex]current bounding box.center)},vertex/.style={anchor=base, circle, minimum size=50mm, inner sep=0pt}]
\path (0.2, 0.9) node[circle, draw, fill=green!30] (s1) {$\sigma_{1}$} (-0.2, -0.9) node[circle, draw, fill=green!30](s2) {$\sigma_{2}$};
\draw[-,black]   (-0.2, -0.5) -- (-0.2, -0.25);
\draw[-,black]   (1.0, 0.25) -- (1.0, 2.0);
\draw[-,black]   (-0.2, -0.25) -- (1.0, 0.25);
\draw[-,black]   (s2) -- (-0.2, -2);
\path (0,0.0) node () {} (-0.95, -0.9) node (s4) { ${\color{blue}{a}}$ };
\path (0,0.0) node () {} (0.55, -0.9) node (s4) { ${\color{red}{b}}$ };
\path (0,0.0) node () {} (0.75,1.7) node (s4) { ${\color{black}{\theta_{1}}}$ };
\path (0,0.0) node () {} (0.15,0.15) node (s4) { ${\color{black}{\theta_{2}}}$ };
\path (0,0.0) node () {} (0.45-0.4,-1.6) node (s4) { ${\color{black}{\theta_{3}}}$ };
\shade [left color=white, right color=gray]
(-2.2+0.6,-2.0) rectangle (-2.0+0.6, 2.0);
\end{tikzpicture}
\qquad
= 2\pi \im \langle \sigma_{1} \rangle_{a} \Delta\langle\sigma_{2}\rangle \frac{ \delta(\theta_{12})  }{ \theta_{23} } \\
\mathfrak{M}^{\sigma_{2}, {\rm disc}} & = \quad 
\begin{tikzpicture}[baseline={([yshift=-.6ex]current bounding box.center)},vertex/.style={anchor=base, circle, minimum size=50mm, inner sep=0pt}]
\path (0.2, 0.9) node[circle, draw, fill=green!30] (s1) {$\sigma_{1}$} (-0.6, -0.9) node[circle, draw, fill=green!30](s2) {$\sigma_{2}$};
\draw[-,black]   (0.2, 1.3) -- (0.2, 2.0);
\draw[-,black]   (0.2, -2.0) -- (0.2, 0.5);
\path (0,0.0) node () {} (-0.55, 0.9) node (s4) { ${\color{blue}{a}}$ };
\path (0,0.0) node () {} (0.95, 0.9) node (s4) { ${\color{red}{b}}$ };
\path (0,0.0) node () {} (0.55, 1.7) node (s4) { ${\color{black}{\theta_{1}}}$ };
\path (0,0.0) node () {} (0.55, 0.0) node (s4) { ${\color{black}{\theta_{2}}}$ };
\path (0,0.0) node () {} (0.55,-1.6) node (s4) { ${\color{black}{\theta_{3}}}$ };
\shade [left color=white, right color=gray]
(-2.2+0.6,-2.0) rectangle (-2.0+0.6, 2.0);
\end{tikzpicture}
\qquad
= 2\pi \im \langle \sigma_{2} \rangle_{a} \Delta\langle\sigma_{1}\rangle \frac{ \delta(\theta_{23})  }{ \theta_{12} } \, .
\end{aligned}
\end{equation}

The calculation of the above diagrams retraces the same arguments already followed for the one point function; thus,
\begin{equation}
\begin{aligned}
\label{}
\{ \mathfrak{M}^{\sigma_{1}, {\rm disc}} \} & = - \langle \sigma_{1} \rangle_{a} \Delta \langle \sigma_{2} \rangle \mathcal{D}(\chi_{2}) + \Os(R^{-1/2}) \, , \\
\{ \mathfrak{M}^{\sigma_{2}, {\rm disc}} \} & = - \langle \sigma_{2} \rangle_{a} \Delta \langle \sigma_{1} \rangle \mathcal{D}(\chi_{1}) + \Os(R^{-1/2}) \, .
\end{aligned}
\end{equation}
Collecting all the results obtained so far and expressing the scaling function $\mathcal{D}(\chi_{j})$ as an integral involving the passage probability, with the aid of (\ref{droplet_profile}), we find
\begin{equation}
\begin{aligned}
\label{result}
\langle \sigma_{1}(x_{1},y_{1}) \sigma_{2}(x_{2},y_{2}) \rangle_{ B_{bab} } & = \langle \sigma_{1} \rangle_{a} \langle \sigma_{2} \rangle_{a} + \Delta\langle\sigma_{1}\rangle  \Delta\langle\sigma_{2}\rangle \int_{0}^{x_{1}}\textrm{d}u_{1} \int_{0}^{x_{2}}\textrm{d}u_{2} P_{2}(u_{1},y_{1};u_{2},y_{2}) \\
& - \langle \sigma_{2} \rangle_{a} \Delta \langle \sigma_{1} \rangle \int_{0}^{x_{1}}\textrm{d}u_{1} \, P_{1}(u_{1},y_{1}) - \langle \sigma_{1} \rangle_{a} \Delta \langle \sigma_{2} \rangle \int_{0}^{x_{2}}\textrm{d}u_{2} \, P_{1}(u_{2},y_{2}) \\
& + \Os(R^{-1/2}) \, .
\end{aligned}
\end{equation}
The first term in the right hand side of (\ref{result}) follows by imposing the correct boundary conditions for $x_{j} \rightarrow + \infty$. Notice also that upon taking one of the two spins deep inside the bulk, the pair correlation function (\ref{result}) satisfies the clustering property
\begin{equation}
\label{cluster1}
\lim_{ x_{2}\rightarrow+\infty } \langle \sigma_{1}(x_{1},y_{1}) \sigma_{2}(x_{2},y_{2}) \rangle_{ B_{bab} } = \langle \sigma_{2} \rangle_{b} \langle \sigma_{1}(x_{1},y_{1}) \rangle_{B_{bab}} \, ,
\end{equation}
and analogously when $x_{1} \rightarrow + \infty$ with finite $x_{2}$.

\begin{figure}[htbp]
\centering
\includegraphics[width=10.5cm]{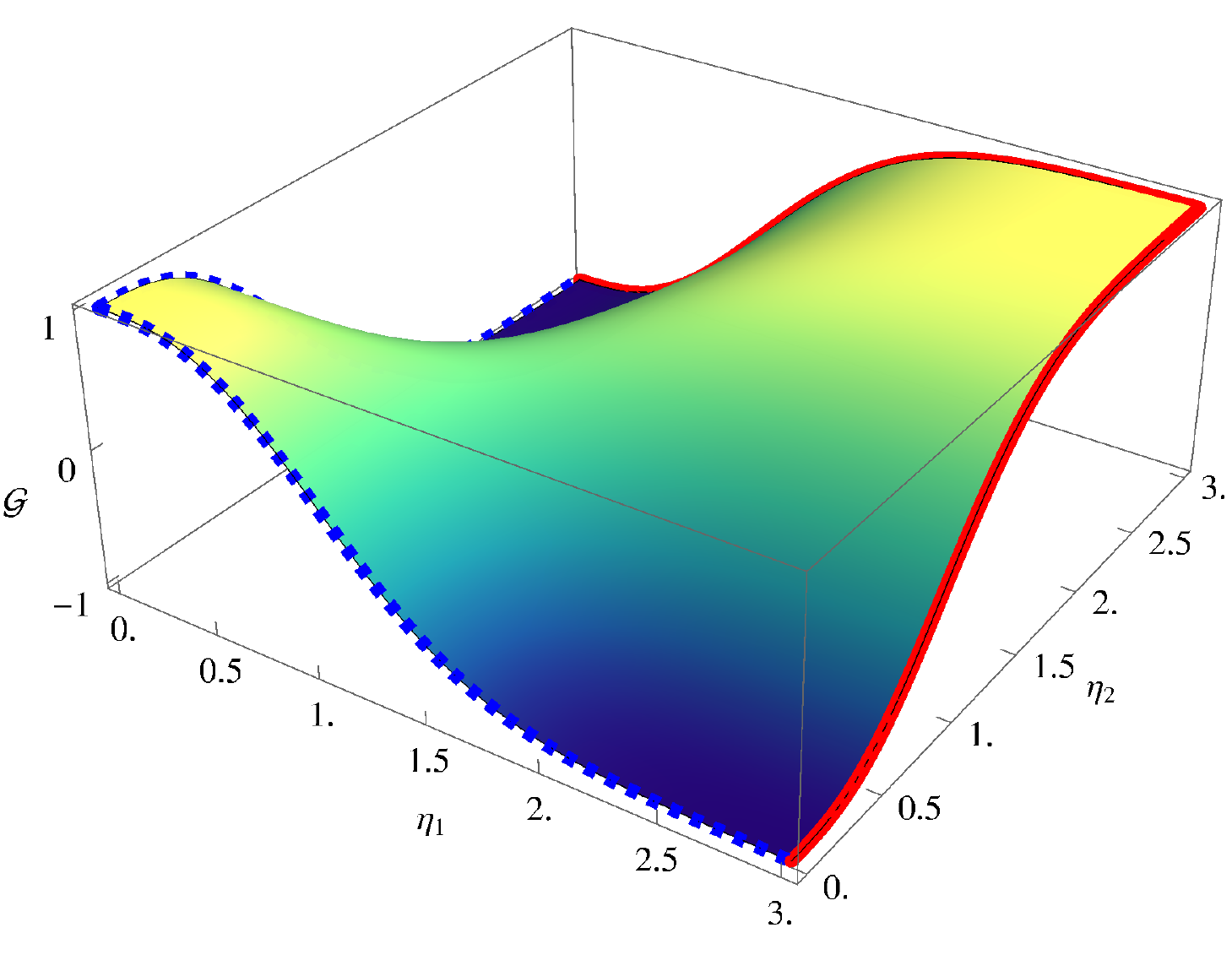}
\caption{The function $\mathcal{G}(\eta_{1},\tau_{1};\eta_{2},\tau_{2})$ with $\tau_{1}=-\tau_{2}=0.1$. The solid red curves and the dashed blue curves correspond to the droplet profiles $\pm\Upsilon(\chi_{1})$ and $\pm\Upsilon(\chi_{2})$, respectively.}
\label{fig_g3}
\end{figure}

In view of future use, we write (\ref{result}) in the following form
\begin{equation}
\begin{aligned}
\label{02092020_1436}
\langle \sigma_{1}(x_{1},y_{1}) \sigma_{2}(x_{2},y_{2}) \rangle_{ B_{bab} } & = \widehat{\langle\sigma_{1}\rangle} \widehat{\langle\sigma_{2}\rangle} \mathcal{G}(\eta_{1},\tau_{1};\eta_{2},\tau_{2}) - \widetilde{\langle\sigma_{1}\rangle} \widehat{\langle\sigma_{2}\rangle} \Upsilon(\chi_{2}) - \widetilde{\langle\sigma_{2}\rangle} \widehat{\langle\sigma_{1}\rangle} \Upsilon(\chi_{1}) \\
& + \widetilde{\langle\sigma_{1}\rangle} \widetilde{\langle\sigma_{2}\rangle} + \Os(R^{-1/2}) \, ,
\end{aligned}
\end{equation}
where $\widetilde{\langle\sigma_{j}\rangle} = (\langle \sigma_{j} \rangle_{a} + \langle \sigma_{j} \rangle_{b})/2$ is the averaged vacuum expectation value in pure phases, $\widehat{\langle\sigma_{j}\rangle} = (\langle \sigma_{j} \rangle_{a} - \langle \sigma_{j} \rangle_{b})/2$ is the half jump of order parameter across the interface, and $\mathcal{G}(\eta_{1},\tau_{1};\eta_{2},\tau_{2})$ is the scaling function which encodes the connected part of the two-point correlation function, i.e.,
\begin{equation}
\begin{aligned}
\label{02092020_1426}
\langle \sigma_{1}(x_{1},y_{1}) \sigma_{2}(x_{2},y_{2}) \rangle_{ B_{bab} }^{\rm CP} & = \widehat{\langle\sigma_{1}\rangle} \widehat{\langle\sigma_{2}\rangle} \mathcal{G}(\eta_{1},\tau_{1};\eta_{2},\tau_{2}) + \Os(R^{-1/2}) \, ,
\end{aligned}
\end{equation}
with
\begin{equation}
\begin{aligned}
\label{02092020_1426b}
\mathcal{G}(\eta_{1},\tau_{1};\eta_{2},\tau_{2}) & = \int_{0}^{\infty}\textrm{d}u_{1} \int_{0}^{\infty}\textrm{d}u_{2} \, P_{2}(u_{1},y_{1};u_{2},y_{2}) \textrm{sign}(x_{1}-u_{1}) \textrm{sign}(x_{2}-u_{2}) \, .
\end{aligned}
\end{equation}
The function $\mathcal{G}$ defined by (\ref{02092020_1426b}) is plotted in Fig.~\ref{fig_g3} as function of the rescaled coordinates $\eta_{1}$ and $\eta_{2}$ with fixed vertical separation $\tau_{1}-\tau_{2}$ of spin fields.
From (\ref{02092020_1426b}) it is simple to establish the following clustering properties:
\begin{equation}
\begin{aligned}
\label{cluster2}
\lim_{x_{1}\rightarrow+\infty} \mathcal{G}(\eta_{1},\tau_{1};\eta_{2},\tau_{2}) & = \Upsilon(\chi_{2}) \, , \\
\lim_{x_{2}\rightarrow+\infty} \mathcal{G}(\eta_{1},\tau_{1};\eta_{2},\tau_{2}) & = \Upsilon(\chi_{1}) \, ,
\end{aligned}
\end{equation}
whose occurrence is visualized by means of the solid red curves at the boundaries of the surface depicted in Fig.~\ref{fig_g3}. In a completely analogous way, one finds
\begin{equation}
\begin{aligned}
\label{cluster3}
\lim_{x_{1}\rightarrow 0} \mathcal{G}(\eta_{1},\tau_{1};\eta_{2},\tau_{2}) & = - \Upsilon(\chi_{2}) \, , \\
\lim_{x_{2}\rightarrow 0} \mathcal{G}(\eta_{1},\tau_{1};\eta_{2},\tau_{2}) & = - \Upsilon(\chi_{1}) \, ,
\end{aligned}
\end{equation}
which correspond to the dashed blue curves visualized in Fig.~\ref{fig_g3}.

A rather more transparent expression for the spin-spin correlation function can be obtained for spin fields arranged parallel to the interface, or, more precisely, parallel to the line which joins the pinning points. For this configuration, $x \equiv x_{1}=x_{2}$, $y \equiv y_{1}=-y_{2}$. By further restricting the focus to the regime of small vertical separations between spin fields, we obtain (\ref{12022021_1651}), which we report here for convenience
\begin{equation}
\begin{aligned}
\label{04092020_1629}
\langle \sigma(x,y) \sigma(x,-y) \rangle_{B_{bab}} & = \left( \frac{ \langle \sigma \rangle_{a} + \langle \sigma \rangle_{b} }{ 2 } \right)^{2} - \frac{ \langle \sigma \rangle_{a}^{2} - \langle \sigma \rangle_{b}^{2} }{ 2 } \Upsilon(\eta) + \\
& + \left( \frac{ \langle \sigma \rangle_{a} - \langle \sigma \rangle_{b} }{ 2 } \right)^{2} \biggl[ 1 - \frac{16}{\pi} \sqrt{\tau} \eta^{2} \textrm{e}^{-\eta^{2}} \biggr] + \Os(\tau^{3/2}) \, ,
\end{aligned}
\end{equation}
for $\xi_{\rm b} \ll y \ll R$. The derivation of (\ref{04092020_1629}) is supplied in Appendix \ref{appD}.

The expression for the parallel correlation function given by (\ref{04092020_1629}) is one of the most important results of this paper. From the result (\ref{04092020_1629}) it is immediate to appreciate how phase separation on the half plane generates long-range correlations. The term proportional to $\sqrt{\tau}$ is the signature of the long-range character of interfacial correlations. Quite interestingly, the term $\sqrt{\tau}$ is multiplied by a function which is proportional to the passage probability, $\eta^{2} \textrm{e}^{-\eta^{2}}$. The entropic factor $\eta^{2}$ suppresses the correlations in the closeness of the wall, while the exponential factor suppresses interfacial correlations in the bulk. The maximum effect is thus achieved when the passage probability reaches its maximum value. The latter is reached for $\eta=1$, corresponding to a distance $x = \sqrt{R\xi_{\rm b}}$ from the wall. This feature -- which emerges from an exact field-theoretic calculation -- could have been guessed from the very beginning once the Brownian excursion character of the interface is established by the calculation of the one-point correlation function.

We conclude this section by observing how the spin-spin correlation function given by (\ref{result}) coincides with the following expression obtained within the probabilistic interpretation
\begin{equation}
\begin{aligned}
\label{cfppr}
\langle \sigma_{1}(x_{1},y_{1}) \sigma_{2}(x_{2},y_{2}) \rangle_{ B_{bab} } & = \int_{0}^{\infty}\textrm{d}u_{1} \int_{0}^{\infty}\textrm{d}u_{2} P_{2}(u_{1},y_{1};u_{2},y_{2}) \sigma_{ab}( x_{1} \vert u_{1} ) \sigma_{ab}( x_{2} \vert u_{2} ) \, ,
\end{aligned}
\end{equation}
where $P_{2}$ is the joint passage probability density (\ref{38}) and $\sigma_{ab}( x_{i} \vert u_{i} )$ is the conditioned magnetization profile given by (\ref{sharp}). If we specialize the above result to the Ising model ($\langle\sigma\rangle_{a/b} = \mp M$), then (\ref{cfppr}) reduces to the corresponding result obtained within the framework of Solid-On-Solid models\footnote{See Eq. (22) of \cite{Burkhardt}.}, as consistency requires. Moreover, for the Ising model the spin-spin correlation function (\ref{02092020_1436}) simplifies to
\begin{equation}
\begin{aligned}
\label{02092020_1429}
\langle \sigma(x_{1},y_{1}) \sigma(x_{2},y_{2}) \rangle_{ B_{+-+} } & = M^{2} \mathcal{G}(\eta_{1},\tau_{1};\eta_{2},\tau_{2}) + \Os(R^{-1/2}) \, .
\end{aligned}
\end{equation}
The subleading correction of order $\Os(R^{-1/2})$ is computed in the next section.

\subsection{Spin-spin correlation function at order $\Os(R^{-1/2})$}
In order to compute the next-to-leading term appearing in (\ref{02092020_1429}), we need to further expand both the bulk and boundary matrix elements at low energies. The connected matrix element comprises the following terms
\begin{eqnarray}
\mathfrak{M}_{D}^{\sigma, {\rm CP}} & = & F_{aba}^{\sigma_{1}}(\theta_{12}+\im \pi) F_{aba}^{\sigma_{2}}(\theta_{23}+\im \pi) \, , \\
\mathfrak{M}_{R}^{\sigma, {\rm CP}} & = & F_{aba}^{\sigma_{1}}(\widehat{\theta}_{12}+\im \pi) \mathcal{R}_{aba}(\theta_{2}) F_{aba}^{\sigma_{2}}(\theta_{23}+\im \pi) \, ,
\end{eqnarray}
while the disconnected ones are given by
\begin{eqnarray}
\mathfrak{M}^{\sigma_{1}, {\rm disc}} & = & 2\pi \langle \sigma_{1} \rangle_{a} F_{aba}^{\sigma_{2}}(\theta_{23}+\im \pi) \delta(\theta_{12}) \, , \\
\mathfrak{M}^{\sigma_{2}, {\rm disc}} & = & 2\pi \langle \sigma_{2} \rangle_{a} F_{aba}^{\sigma_{1}}(\theta_{12}+\im \pi) \delta(\theta_{23}) \, .
\end{eqnarray}
The form factor $F_{aba}^{\sigma_{l}}(\theta_{ij}+\im \pi)$ is expanded as in (\ref{lowthetaFF}) but now the expansion coefficients are $c_{k}^{(l)}$ with an extra superscript $l$ which labels the spin field. In an analogous way, we proceed by expanding the boundary $S$-matrix at low rapidities; hence, we write
\begin{equation}
\label{ }
\mathcal{R}_{aba}(\theta) = -1 + \im \mathfrak{r} \, \theta + \Os(\theta^{2}) \, ,
\end{equation}
where $\mathfrak{r}$ is a model-dependent (real) coefficient which is known for integrable field theories \cite{LS, BH_2016}. The connected part of the correlation function is written as follows
\begin{equation}
\begin{aligned}
\label{}
\langle \sigma_{1}(x_{1},y_{1}) \sigma_{2}(x_{2},y_{2}) \rangle_{ B_{bab} }^{\rm CP} & = \frac{1}{\mathcal{Z}} \int \frac{\textrm{d}\theta_{1}\textrm{d}\theta_{2}\textrm{d}\theta_{3}}{(2\pi)^{3}} \biggl[ \mathcal{F}_{\mu}(\theta_{1}) \mathfrak{M}_{D}^{\sigma, {\rm CP}} \mathcal{F}_{\mu}^{*}(\theta_{3}) U_{+} + \mathcal{F}_{\mu}(\theta_{1}) \mathfrak{M}_{D}^{\sigma, {\rm CP}} \mathcal{F}_{\mu}^{*}(\theta_{3}) U_{-} \biggr] \, .
\end{aligned}
\end{equation}
The low-rapidity expansion of the ``directed'' matrix element yields
\begin{equation}
\label{1545a}
\mathcal{F}_{\mu}(\theta_{1}) \mathfrak{M}_{D}^{\sigma, {\rm CP}} \mathcal{F}_{\mu}^{*}(\theta_{3}) U_{+} = \mathfrak{a}^{2}c_{-1}^{(1)}c_{-1}^{(2)} \frac{\theta_{1}\theta_{3}}{\theta_{12}\theta_{23}}  \biggl[ 1 - \im \omega_{1} \theta_{12} - \im \omega_{2} \theta_{23} \biggr] + \Os(\theta^{2}) \, ,
\end{equation}
where $\omega_{l}=\mathfrak{b}/\mathfrak{a}+\im c_{0}^{(l)}/c_{-1}^{(l)}$. The ``reflected'' matrix element gives
\begin{equation}
\label{1545}
\mathcal{F}_{\mu}(\theta_{1}) \mathfrak{M}_{R}^{\sigma, {\rm CP}} \mathcal{F}_{\mu}^{*}(\theta_{3}) U_{+} = - \mathfrak{a}^{2}c_{-1}^{(1)}c_{-1}^{(2)} \frac{\theta_{1}\theta_{3}}{\widehat{\theta}_{12}\theta_{23}}  \biggl[ 1 - \im \omega_{1} \widehat{\theta}_{12} - \im \omega_{2} \theta_{23} + \im \omega^{\prime} \theta_{2} \biggr] + \Os(\theta^{2}) \, ,
\end{equation}
with $\omega^{\prime} = -\mathfrak{r} + 2\mathfrak{b}/ \mathfrak{a}$. Since the boundary form factor $\mathcal{F}_{\mu}(\theta)$ and the boundary $S$-matrix $\mathcal{R}_{aba}(\theta)$ are related by means of the functional equation $\mathcal{F}_{\mu}(\theta)=\mathcal{R}_{aba}(\theta)\mathcal{F}_{\mu}(-\theta)$ \cite{BPT_BFF}, it follows that $2\mathfrak{b}=\mathfrak{a}\mathfrak{r}$ and therefore $\omega^{\prime}=0$. As a result, this crucial observation removes the asymmetry between (\ref{1545a}) and (\ref{1545}).

In general, also the partition function has to be expanded in an analogous way. However, it turns out that for $\mathcal{Z}$ it is legitimate to use the expression (\ref{e005}) because the large-$R$ expansion involves higher-order powers of $R^{-1/2}$. By using the ``bag''-notation given in (\ref{02092020_1057}), the correlation function including corrections proportional to $R^{-1/2}$ reads
\begin{equation}
\begin{aligned}
\label{1546}
\langle \sigma_{1}(x_{1},y_{1}) \sigma_{2}(x_{2},y_{2}) \rangle_{ B_{bab} }^{\rm CP} & = \frac{1}{\pi^{5/2}} c_{-1}^{(1)} c_{-1}^{(2)} \biggl[ \left( \Lbag \frac{\theta_{1}\theta_{3}}{\widehat{\theta}_{12}\theta_{23}} \Rbag_{+} - \Lbag \frac{\theta_{1}\theta_{3}}{\theta_{12}\theta_{23}} \Rbag_{-} \right) + \\
& - \im \omega_{1} \sqrt{\frac{2}{mR}} \left( \Lbag \frac{\theta_{1}\theta_{3}}{\theta_{23}} \Rbag_{+} - \Lbag \frac{\theta_{1}\theta_{3}}{\theta_{23}} \Rbag_{-} \right) - \im \omega_{2} \sqrt{\frac{2}{mR}} \left( \Lbag \frac{\theta_{1}\theta_{3}}{\theta_{12}} \Rbag_{+} - \Lbag \frac{\theta_{1}\theta_{3}}{\widehat{\theta}_{12}} \Rbag_{-} \right) \biggr] \\
& + \Os(R^{-1}) \, .
\end{aligned}
\end{equation}
The terms at order $\Os(\theta^{2})$ in (\ref{1545a}) and (\ref{1545}) contribute of order $R^{-1}$ in (\ref{1546}). The first term in the right hand side of (\ref{1546}) originates the leading order contribution, while the remaining two terms give corrections at order $R^{-1/2}$. In the following, we outline the calculation of these corrections. By taking the first derivative with respect to $\eta_{2}$ in the first term, we find
\begin{equation}
\begin{aligned}
\label{}
\im \partial_{\eta_{2}} \left( \Lbag \frac{\theta_{1}\theta_{3}}{\theta_{23}} \Rbag_{+} - \Lbag \frac{\theta_{1}\theta_{3}}{\theta_{23}} \Rbag_{-} \right) & = \im^{2} \left( \Lbag \theta_{1}\theta_{3} \Rbag_{+} - \Lbag \theta_{1}\theta_{3} \Rbag_{-} \right) \\
& = \pi^{5/2} \im^{2} \lambda^{2} P_{2}(x_{1},y_{1};x_{2},y_{2}) \, ,
\end{aligned}
\end{equation}
integrating back and imposing the boundary conditions, we readily obtain
\begin{equation}
\label{ }
\im \left( \Lbag \frac{\theta_{1}\theta_{3}}{\theta_{23}} \Rbag_{+} - \Lbag \frac{\theta_{1}\theta_{3}}{\theta_{23}} \Rbag_{-} \right) = - \pi^{5/2} \lambda \int_{0}^{x_{2}}\textrm{d}u_{2} P_{2}(x_{1},y_{1};u_{2},y_{2}) \, .
\end{equation}
Proceeding in an analogous way for the other term and collecting all the results together, the connected part reads
\begin{equation}
\begin{aligned}
\label{}
\langle \sigma_{1}(x_{1},y_{1}) \sigma_{2}(x_{2},y_{2}) \rangle_{ B_{bab} }^{\rm CP} & = \frac{\Delta \langle \sigma_{1} \rangle\Delta \langle \sigma_{2} \rangle}{4} \mathcal{G}(\eta_{1},\tau_{1};\eta_{2},\tau_{2}) \\
& - \frac{\omega_{1}}{m} \Delta \langle \sigma_{1} \rangle \Delta \langle \sigma_{2} \rangle \int_{0}^{x_{2}}\textrm{d}u_{2} P_{2}(x_{1},y_{1};u_{2},y_{2}) \\
& - \frac{\omega_{2}}{m} \Delta \langle \sigma_{1} \rangle \Delta \langle \sigma_{2} \rangle \int_{0}^{x_{1}}\textrm{d}u_{1} P_{2}(u_{1},y_{1};x_{2},y_{2}) + \Os(R^{-1}) \, .
\end{aligned}
\end{equation}
The disconnected terms are proportional to the one-point correlation functions (\ref{e011}); thus,
\begin{equation}
\begin{aligned}
\label{}
\{ \mathfrak{M}_{1}^{\rm disc} \} & = - \langle \sigma_{1} \rangle_{a} \Delta \langle \sigma_{2} \rangle \int_{0}^{x_{2}}\textrm{d}u_{2} \, P_{1}(u_{2},y_{2}) + \omega_{2} \langle \sigma_{1} \rangle_{a} \Delta \langle \sigma_{2} \rangle P_{1}(x_{2},y_{2})/m + \Os(R^{-1}) \, , \\
\{ \mathfrak{M}_{2}^{\rm disc} \} & = - \langle \sigma_{2} \rangle_{a} \Delta \langle \sigma_{1} \rangle \int_{0}^{x_{1}}\textrm{d}u_{1} \, P_{1}(u_{1},y_{1}) + \omega_{1} \langle \sigma_{2} \rangle_{a} \Delta \langle \sigma_{2} \rangle P_{1}(x_{1},y_{1})/m + \Os(R^{-1}) \, .
\end{aligned}
\end{equation}
We can cast the large-$R$ expansion of the correlation function in the form
\begin{equation}
\label{090738}
\langle \sigma_{1}(x_{1},y_{1}) \sigma_{2}(x_{2},y_{2}) \rangle_{ B_{bab} } = \sum_{\ell=0}^{\infty} \Bigl[ \langle \sigma_{1}(x_{1},y_{1}) \sigma_{2}(x_{2},y_{2}) \rangle_{ B_{bab} } \Bigr]_{\ell} \, ,
\end{equation}
where $[ \dots ]_{\ell} = \Os(R^{-\ell/2})$. Therefore the term with $\ell=0$, which is the leading one, is given by (\ref{02092020_1436}). The first subleading correction, which is the term with $\ell=1$, is given by
\begin{equation}
\begin{aligned}
\label{}
\Bigl[ \langle \sigma_{1}(x_{1},y_{1}) \sigma_{2}(x_{2},y_{2}) \rangle_{ B_{bab} } \Bigr]_{1} & = \frac{\omega_{1}\Delta \langle \sigma_{1}\rangle}{m} \biggl[ \langle \sigma_{2} \rangle_{a} P_{1}(x_{1},y_{1}) - \Delta \langle \sigma_{2} \rangle \int_{0}^{x_{2}}\textrm{d}u_{2} P_{2}(x_{1},y_{1};u_{2},y_{2}) \biggr] + \\
& + \frac{\omega_{2}\Delta \langle \sigma_{2}\rangle}{m} \biggl[ \langle \sigma_{1} \rangle_{a} P_{1}(x_{2},y_{2}) - \Delta \langle \sigma_{1} \rangle \int_{0}^{x_{1}}\textrm{d}u_{1} P_{2}(u_{1},y_{1};x_{2},y_{2}) \biggr] \, .
\end{aligned}
\end{equation}
The clustering property (\ref{cluster1}), which we have checked at order $\ell=0$ in (\ref{cluster2}), is satisfied also at order $\ell=1$. Taking $x_{2}$ deep into the $b$ phase the vacuum expectation value $\langle \sigma_{2} \rangle_{b}$ factors out and one is left with the term at order $R^{-1/2}$ stemming from the one-point function in the variable $x_{1}$, i.e.
\begin{equation}
\label{cluster5}
\lim_{x_{2}\rightarrow+\infty} \Bigl[ \langle \sigma_{1}(x_{1},y_{1}) \sigma_{2}(x_{2},y_{2}) \rangle_{ B_{bab} } \Bigr]_{1} =  \langle \sigma_{2} \rangle_{b} \biggl[ \frac{\omega_{1}\Delta \langle \sigma_{1}\rangle}{m} P_{1}(x_{1},y_{1}) \biggr] \, ,
\end{equation}
as consistency requires. 

To conclude this section, we present an explicit evaluation of the subleading correction for the Ising model. To be definite, we examine the parallel correlation function for which the interface structure correction reads
\begin{equation}
\begin{aligned}
\label{sub_Ising_model}
\Bigl[ \langle \sigma_{1}(x,y) \sigma_{2}(x,-y) \rangle_{ B_{+-+} } \Bigr]_{1} & = \frac{2M^{2}}{m} \biggl[ -P_{1}(x,y) + 2 \int_{0}^{x}\textrm{d}u \, P_{2}(x,y;u,-y) \biggr] \\
& = \frac{M^{2}}{\sqrt{2mR}} \mathcal{B}_{\parallel}(\eta,\tau) \, ,
\end{aligned}
\end{equation}
with
\begin{equation}
\mathcal{B}_{\parallel}(\eta,\tau) = \frac{16}{\sqrt{\pi}\kappa} \chi^{2} \textrm{e}^{-\chi^{2}} \biggl[ \textrm{erf}(\chi\sqrt{\tau}) + \textrm{erf}(\chi/\sqrt{\tau}) - 1 \biggr] - \frac{32}{\pi\kappa} \frac{\chi\sqrt{\tau}}{1-\tau} \textrm{e}^{-\chi^{2}} \biggl[ \textrm{e}^{-\chi^{2}\tau} - \textrm{e}^{-\chi^{2}/\tau} \biggr] \, ;
\end{equation}
we recall that $\chi=\eta/\sqrt{1-\tau^{2}}$. The function $\mathcal{B}_{\parallel}(\eta,\tau)$ satisfies the following properties: $\mathcal{B}_{\parallel}(0,\tau)=0$ and $\mathcal{B}_{\parallel}(\eta \rightarrow +\infty,\tau)=0$.

\subsection{Parallel and perpendicular correlation functions}
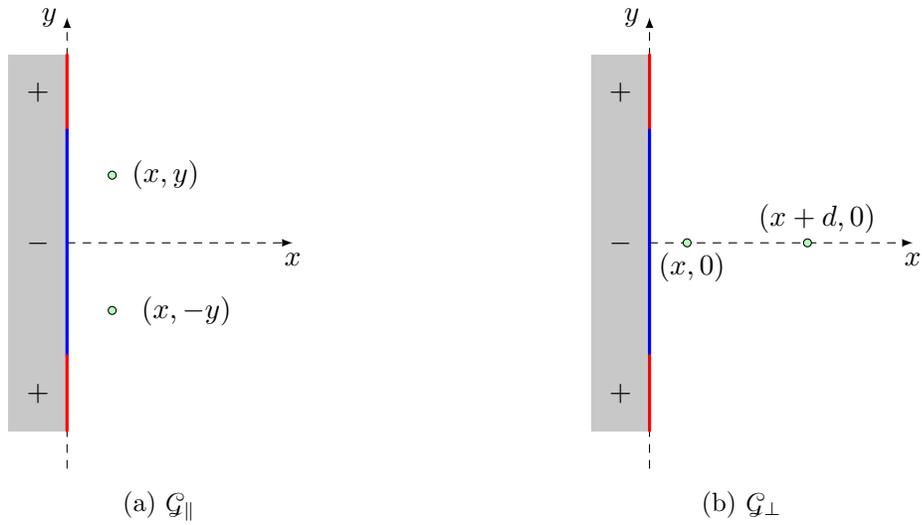
\begin{figure*}[htbp]
\centering
\hspace{20mm}
        \begin{subfigure}[b]{0.25\textwidth}
            \centering
\begin{tikzpicture}[thick, line cap=round, >=latex, scale=0.5]
\tikzset{fontscale/.style = {font=\relsize{#1}}}
\shade [left color=mygray, right color=mygray]
(-1.55,-5.0) rectangle (0, 5.0);
\draw[thin, dashed, ->] (0.0, 0.0) -- (6.0, 0.0) node[below] {$x$};
\draw[thin, dashed, ->] (0.0, -6.0) -- (0.0, 6.0) node[left] {$y$};
\draw[very thick, red, -] (0, 3) -- (0,5) node[] {};
\draw[very thick, blue, -] (0, -3) -- (0,3) node[] {};
\draw[very thick, red, -] (0, -3) -- (0,-5) node[] {};
\draw[thin, fill=white] (-0.2, 0.0) circle (0pt) node[left] {${\color{black}{-}}$};;
\draw[thin, fill=white] (-0.2, 4.0) circle (0pt) node[left] {${\color{black}{+}}$};;
\draw[thin, fill=white] (-0.2, -4.0) circle (0pt) node[left] {${\color{black}{+}}$};;
\draw[thin, fill=green!30] (1.2, 1.8) circle (3pt) node[right] {\,\,$(x,y)$};;
\draw[thin, fill=green!30] (1.2, -1.8) circle (3pt) node[right] {\,\, $(x,-y)$};;
\end{tikzpicture}
            \caption[]%
            {{\small $\mathcal{G}_{\parallel}$}}    
            \label{fig7a}
        \end{subfigure}
\hfill
        \begin{subfigure}[b]{0.25\textwidth}
            \centering
\begin{tikzpicture}[thick, line cap=round, >=latex, scale=0.5]
\tikzset{fontscale/.style = {font=\relsize{#1}}}
\shade [left color=mygray, right color=mygray]
(-1.55,-5.0) rectangle (0, 5.0);
\draw[thin, dashed, ->] (0.0, 0.0) -- (7.0, 0.0) node[below] {$x$};
\draw[thin, dashed, ->] (0.0, -6.0) -- (0.0, 6.0) node[left] {$y$};
\draw[very thick, red, -] (0, 3) -- (0,5) node[] {};
\draw[very thick, blue, -] (0, -3) -- (0,3) node[] {};
\draw[very thick, red, -] (0, -3) -- (0,-5) node[] {};
\draw[thin, fill=white] (-0.2, 0.0) circle (0pt) node[left] {${\color{black}{-}}$};;
\draw[thin, fill=white] (-0.2, 4.0) circle (0pt) node[left] {${\color{black}{+}}$};;
\draw[thin, fill=white] (-0.2, -4.0) circle (0pt) node[left] {${\color{black}{+}}$};;
\draw[thin, fill=green!30] (1.0, 0.0) circle (3pt) node[below] {\,\,$(x,0)$};;
\draw[thin, fill=green!30] (4.2, 0.0) circle (3pt) node[above] {\,\, $(x+d,0)$};;
\end{tikzpicture}
            \caption[]%
            {{\small $\mathcal{G}_{\bot}$}}    
            \label{fig7b}
        \end{subfigure}
\hspace{20mm}
\vskip\baselineskip
\caption[]
{\small Definition of parallel (a) and perpendicular (b) correlation function for the Ising model.}
\label{fig_pa_pe}
\end{figure*}

We specialize the general result (\ref{02092020_1436}) to the two-dimensional Ising model and focus on particularly symmetric configurations for the two spin fields. We have already considered the so-called parallel correlation function ($\parallel$). Here, we also introduce the perpendicular correlation function ($\bot$), meaning that spin fields are arranged perpendicularly to the line which joins the pinning points. The aforementioned correlation functions are defined by
\begin{equation}
\begin{aligned}
\label{02092020_2025}
\mathcal{G}_{\parallel}(\eta,\tau) & = \langle \sigma(x,y) \sigma(x,-y) \rangle_{B_{+-+}}/M^{2}  \, , \\
\mathcal{G}_{\bot}(\eta,\delta) & = \langle \sigma(x,0) \sigma(x+d,0) \rangle_{B_{+-+}}/M^{2} \, , \qquad \delta=d/\lambda \, ,
\end{aligned}
\end{equation}
with $\eta=x/\lambda$ and $\tau=2y/R$. The arrangement of spin fields defining the correlation functions given in (\ref{02092020_2025}) is illustrated in Fig.~\ref{fig_pa_pe}.

\begin{figure}[htbp]
\centering
\includegraphics[width=11.5cm]{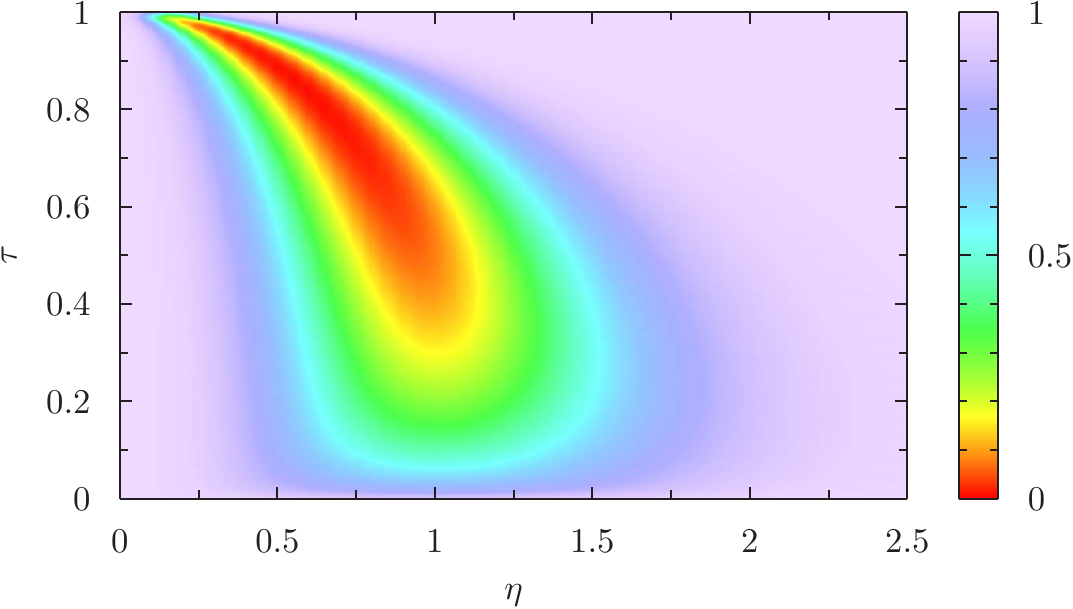}
\caption{The parallel function at leading order: $\mathcal{G}(\eta,\tau;\eta,-\tau)$.}
\label{fig_density_gparallel}
\end{figure}

Thanks to (\ref{02092020_1429}) and (\ref{sub_Ising_model}), the parallel correlation function including interface structure corrections at order $R^{-1/2}$ is given by
\begin{equation}
\begin{aligned}
\label{G_parallel_Ising}
\mathcal{G}_{\parallel}(\eta,\tau) & = \mathcal{G}(\eta,\tau;\eta,-\tau) + \frac{1}{\sqrt{2mR}} \mathcal{B}_{\parallel}(\eta,\tau) + \Os(R^{-1}) \, .
\end{aligned}
\end{equation}
The leading term in the parallel correlation function is shown in Fig.~\ref{fig_density_gparallel} as function of both $\eta$ and $\tau$. As expected, the largest variation occurs within the interfacial region. The latter corresponds to $\eta \approx 1$ when $\tau$ is small and in the closeness of the pinning points when $\tau$ tends to one. These features are visible in the plot of Fig.~\ref{fig_density_gparallel}.

The analysis of the perpendicular correlation function $\mathcal{G}_{\bot}(\eta,\delta)$ requires special care. The reason is due to the fact that spin fields do no longer satisfy the assumption $y_{1}-y_{2} \gg \xi_{\rm b}$. Although the vertical separation vanishes identically, the horizontal separation $d$ is taken to be large compared to the bulk correlation length. In this setup, we can perform the limit $y_{1},y_{2} \rightarrow 0$ in the field-theoretical result with finite $d$. The aforementioned limit actually implies that the correlation coefficient $\rho$ tends to one. Analytic results within the probabilistic description are found by taking $\rho \rightarrow 1$ in the joint passage probability density, which thus reduces to $P_{1}(x_{1},0)\delta(x_{1}-x_{2})$. Therefore when the two spin fields are widely separated from each other ($md \gg 1$) with the leftmost spin field far from the wall ($mx\gg1$) the perpendicular correlation function reads
\begin{equation}
\label{G_perpendicular_Ising}
\mathcal{G}_{\bot}(\eta,\delta) = 1 + \Upsilon(\eta) - \Upsilon(\eta+\delta) + \frac{1}{m} \left( P_{1}(x,0) - P_{1}(x+d,0) \right) + \Os(R^{-1}) \, , \qquad (\delta >0) \, .
\end{equation}
The limit $md \rightarrow \infty$ in (\ref{G_perpendicular_Ising}) yields
\begin{equation}
\label{23012021_1614}
\mathcal{G}_{\bot}(\eta,\delta \rightarrow + \infty) = \Upsilon(\eta) + \frac{P_{1}(x,0)}{m} + \Os(R^{-1}) \, .
\end{equation}
The above result indicates that the clustering to the one-point correlation function is correctly retrieved. All the above features have been accurately verified in \cite{droplet_MC}.


\section{Interface structure factor}
\label{section_6}
Long-range interfacial correlations are traditionally studied in momentum space through the notion of interface structure factor \cite{BKV, MD, PRWE_2014, HD_2015}. The interface structure factor can be obtained upon performing a parallel Fourier transform, i.e., along the direction parallel to the interface, of a suitably defined connected pair correlation function. The interface structure factor is defined by
\begin{equation}
\label{03092020_0904}
\widehat{S}(q) = \frac{ 1 }{ 2 \left( \Delta \langle \sigma \rangle \right)^{2} } \int_{-R/2}^{R/2}\textrm{d}y \, \textrm{e}^{\im q y } \int_{0}^{\infty}\textrm{d}x_{1} \int_{0}^{\infty}\textrm{d}x_{2} \, \langle \sigma(x_{1},y) \sigma(x_{2},-y) \rangle_{B_{bab}}^{\rm conn.} \, ;
\end{equation}
the hat symbol on top of $S(q)$ stands for the presence of finite-size corrections due to the finite distance between pinning points. The connected correlation function $\langle \sigma_{1}(x_{1},y_{1}) \sigma_{2}(x_{2},y_{2}) \rangle_{B_{bab}}^{\rm conn.}$ can be obtained from (\ref{02092020_1436}) by subtracting the disconnected parts which lead to a vanishing correlation function in the limit when $x_{1}$ and/or $x_{2}$ go to infinity. Moreover, since (\ref{03092020_0904}) has to take into account \emph{only} those degrees of freedom coupled to the interface, bulk contributions must be subtracted too. In order to find the appropriate subtraction scheme, we follow the guidelines outlined in \cite{DS_twopoint} by adopting some modifications due to the specificities of the half-plane geometry which induces the droplet-shaped interface. The connected correlation function is thus written as follows
\begin{equation}
\begin{aligned}
\label{03092020_0914}
\langle \sigma_{1}(x_{1},y_{1}) \sigma_{2}(x_{2},y_{2}) \rangle_{B_{bab}}^{\rm conn.} & = \mathscr{G}_{\mathfrak{s}}(x_{1},y_{1};x_{2},y_{2}) - \mathscr{G}_{\rm b}(x_{1},y_{1};x_{2},y_{2}) \, .
\end{aligned}
\end{equation}
The term $\mathscr{G}_{\rm b}(x_{1},y_{1};x_{2},y_{2}) = \langle \sigma_{1} \sigma_{2} \rangle_{b} - \langle \sigma_{1} \rangle_{b} \langle \sigma_{2} \rangle_{b}$ is the connected \emph{bulk} correlation function. The subtraction of the connected bulk correlation function $\mathscr{G}_{\rm b}$ ensures that $\langle \sigma_{1}(x_{1},y_{1}) \sigma_{2}(x_{2},y_{2}) \rangle_{B_{bab}}^{\rm conn.}$ goes to zero when both $x_{1}$ and $x_{2}$ tend to $+\infty$ (deep into the $b$-phase) with their relative distance kept finite. These bulk contributions to the two-point correlation function are visualized within the pictorial representation of matrix elements in (\ref{03092020_1008}).
\begin{equation}
\begin{aligned}
\label{03092020_1008}
\left( \mathcal{M}_{ab}^{\sigma_{1}}\mathcal{M}_{ab}^{\sigma_{2}} \right)^{\rm bulk} \,\, &
=
\,\,
\begin{tikzpicture}[baseline={([yshift=-.6ex]current bounding box.center)},vertex/.style={anchor=base, circle, minimum size=50mm, inner sep=0pt}]
\path (0.2+0.4, 0.9) node[circle, draw, fill=green!30] (s1) {$\sigma_{1}$} (-0.2+0.4, -0.9) node[circle, draw, fill=green!30](s2) {$\sigma_{2}$};
\path (s1) edge [bend left] (s2);
\path (s1) edge [bend right] (s2);
\path (0,0.0) node () {} (-1.1, 0) node (s4) { ${\color{blue}{a}}$ };
\path (0,0.0) node () {} (-1.1+0.75, 0) node (s4) { ${\color{red}{b}}$ };
\shade [left color=white, right color=red]
(-2.2+0.6,1.5) rectangle (-2.0+0.6, 2.0);
\shade [left color=white, right color=blue]
(-2.2+0.6,-1.5) rectangle (-2.0+0.6, 1.5);
\shade [left color=white, right color=red]
(-2.2+0.6,-2.0) rectangle (-2.0+0.6, -1.5);

\draw[black!70]
(-1.4, -1.5) ..controls +(0.0, 0.0) and ( $(-0.7, 0) + (0, -1)$ )..
(-0.7, 0) ..controls +(0.0, 1) and ( $(-1.4, 1.5) + (0, 0)$ )..
(-1.4, 1.5);
\end{tikzpicture}
\qquad
+
\qquad
\begin{tikzpicture}[baseline={([yshift=-.6ex]current bounding box.center)},vertex/.style={anchor=base, circle, minimum size=50mm, inner sep=0pt}]
\path (0.2-0.3, 0.9) node[circle, draw, fill=green!30] (s1) {$\sigma_{1}$} (-0.2-0.3, -0.9) node[circle, draw, fill=green!30](s2) {$\sigma_{2}$};
\path (s1) edge [bend left] (s2);
\path (s1) edge [bend right] (s2);
\path (0,0.0) node () {} (0.7, 0) node (s4) { ${\color{blue}{a}}$ };
\path (0,0.0) node () {} (1.5, 0) node (s4) { ${\color{red}{b}}$ };
\shade [left color=white, right color=red]
(-2.2+0.6,1.5) rectangle (-2.0+0.6, 2.0);
\shade [left color=white, right color=blue]
(-2.2+0.6,-1.5) rectangle (-2.0+0.6, 1.5);
\shade [left color=white, right color=red]
(-2.2+0.6,-2.0) rectangle (-2.0+0.6, -1.5);

\draw[black!70]
(-1.4, -1.5) ..controls +(2.0, 0.0) and ( $(1.17, 0) + (0, -1)$ )..
(1.17, 0) ..controls +(0.0, 1) and ( $(-1.4, 1.5) + (2, 0)$ )..
(-1.4, 1.5);
\end{tikzpicture}
\qquad
+
\qquad
\dots \, .
\end{aligned}
\end{equation}
The two diagrams illustrated in (\ref{03092020_1008}) are obtained upon inserting a multi-kink state in the resolution of the identity between spin fields. When $R$ is sufficiently large both the diagrams yield the bulk correlation function averaged over the two phases separated by the droplet. It is thus clear how the diagrams shown in (\ref{03092020_1008}) involve the propagation of three kinks; as a result, their contribution is definitely subleading and from now on they will be ignored in the discussion that follows.

Let us comment on the first term in the right hand side of (\ref{03092020_0914}). The correlation function $ \mathscr{G}_{\mathfrak{s}}(x_{1},y_{1};x_{2},y_{2})$ is defined by
\begin{equation}
\begin{aligned}
\label{}
\mathscr{G}_{\mathfrak{s}}(x_{1},y_{1};x_{2},y_{2}) & = \bigl\langle \bigl[ \sigma_{1}(x_{1},y_{1}) - \mathfrak{s}_{ab}(x_{1},y_{1}) \bigr] \bigl[ \sigma_{2}(x_{2},y_{2}) - \mathfrak{s}_{ab}(x_{2},y_{2}) \bigr] \bigr\rangle_{B_{bab}} \, ,
\end{aligned}
\end{equation}
with $\mathfrak{s}_{ab}(x_{j},y_{j})$ a reference density profile which tends to the asymptotic value $\langle \sigma_{j} \rangle_{b}$ when $x_{j}$ tends to $+ \infty$. It is evident how the above asymptotic specification does not fix the reference density profile. On phenomenological grounds, a rather natural candidate is provided by the sharp reference profile 
\begin{equation}
\label{084518}
\mathfrak{s}_{ab}^{(\rm sharp)}(x_{j},y_{j};L) = \langle \sigma_{j} \rangle_{a} \theta(L-x_{j}) + \langle \sigma_{j} \rangle_{b} \theta(x_{j}-L) \, ,
\end{equation}
with a parameter $L$ that indicates the position in which the two phases meet within the sharp interface picture. The above can be regarded as the natural extension to the half-plane of the prescription used in \cite{BKV}\footnote{It should be implicit that fluctuations of the contour separating the two phases do not preserve the droplet volume. For the sake of completeness we mention that constraining the droplet volume leads to interesting effects on free energies \cite{RSVDG_2019} and on correlations in interfacial phenomena \cite{Gross_2018}.}. However, we already observe at this stage the occurrence of a specific feature of the half-plane geometry. In fact, while $L=0$ in the absence of the vertical wall -- see \cite{DS_twopoint} for the strip geometry in $d=2$ dimensions and \cite{BKV} for $d=3$ -- for the half-plane $L \sim \sqrt{R}$ and, moreover, $L$ depends also on $y$ through the combination $\kappa$ by means of $L \propto \kappa \sqrt{R}$; see the contour line equation (\ref{15022021_0837}). It is thus evident that the precise form of $L$ is not fixed \emph{a priori} and therefore the definition (\ref{03092020_0914}) with the prescription (\ref{084518}) contains an ad-hoc parameter. 

In order to construct a definition of connected correlation function which is \emph{free} of ad-hoc parameters, we will consider for the reference profile the one-point function at leading order; thus, we set
\begin{equation}
\label{084637}
\mathfrak{s}_{ab}(x_{j},y_{j}) = \bigl[ \langle \sigma_{j}(x_{j},y_{j}) \rangle_{B_{bab}} \bigr]_{0} \, .
\end{equation}
Thanks to the above choice the ambiguity inherent to the accessory parameter $L$ is removed. As a result, (\ref{084637}) provides an intrinsically natural choice free of external parameters. For the sake of completeness, it has to be mentioned how the full magnetization profile with subleading corrections also provides a parameter-free reference profile. Nonetheless, we will adopt (\ref{084637}) as the natural choice for the reference profile and other prescriptions can be examined by properly modifying the treatment that follows. 

Leaving out the subdominant bulk corrections mentioned above, the interface structure factor reads
\begin{equation}
\label{03092020_0940}
\widehat{S}(q) = \frac{ 1 }{ 2 \left( \Delta \langle \sigma \rangle \right)^{2} } \int_{-R/2}^{R/2}\textrm{d}y \, \textrm{e}^{\im q y } \int_{0}^{\infty}\textrm{d}x_{1} \int_{0}^{\infty}\textrm{d}x_{2} \, \mathscr{G}_{\mathfrak{s}}(x_{1},y;x_{2},-y) \, .
\end{equation}
It is convenient to split the reference profile $\mathfrak{s}_{ab}$ as follows
\begin{equation}
\label{}
\mathfrak{s}_{ab}(x_{j},y_{j}) = \langle \sigma_{j}(x_{j},y_{j}) \rangle_{B_{bab}} + \mathfrak{e}_{ab}(x_{j},y_{j}) \, ,
\end{equation}
with the excess part defined by
\begin{equation}
\begin{aligned}
\label{02062021_0830}
\mathfrak{e}_{ab}(x_{j},y_{j}) & = \bigl[ \langle \sigma_{j}(x_{j},y_{j}) \rangle_{B_{bab}} \bigr]_{0} - \langle \sigma_{j}(x_{j},y_{j}) \rangle_{B_{bab}}  \, , \\
& = - A_{ab}^{(\sigma_{j})} P_{1}(x_{j},y_{j}) + \Os(R^{-1}) \, ,
\end{aligned}
\end{equation}
with the structure amplitude
\begin{equation}
\label{ }
A_{ab}^{(\sigma_{j})} = \frac{c_{0}^{(j)}}{m} + \frac{\mathfrak{b}}{\mathfrak{a}} \frac{\Delta\langle\sigma_{j}\rangle}{m} \, ,
\end{equation}
in agreement with (\ref{01092020_2047}). It then follows that
\begin{equation}
\label{ }
\mathscr{G}_{\mathfrak{s}}(x_{1},y_{1};x_{2},y_{2}) = G(x_{1},y_{1};x_{2},y_{2}) + \mathfrak{e}_{ab}(x_{1},y_{1}) \mathfrak{e}_{ab}(x_{2},y_{2})
\end{equation}
with
\begin{equation}
\label{ }
G(x_{1},y_{1};x_{2},y_{2}) = \langle \sigma_{1}(x_{1},y_{1}) \sigma_{2}(x_{2},y_{2}) \rangle_{B_{bab}} - \langle \sigma_{1}(x_{1},y_{1}) \rangle_{B_{bab}} \langle \sigma_{2}(x_{2},y_{2}) \rangle_{B_{bab}} \, .
\end{equation}

The integrand in (\ref{03092020_0940}) is then written as a power series in the small parameter $(mR)^{-1/2}$; thus, we write
\begin{equation}
\label{}
\mathscr{G}_{\mathfrak{s}}(x_{1},y;x_{2},-y) = \sum_{\ell=0}^{\infty} \Bigl[ \mathscr{G}_{\mathfrak{s}}(x_{1},y;x_{2},-y) \Bigr]_{\ell} \, .
\end{equation}
In analogy with (\ref{090738}), the subscript $\ell$ stands for a term of order $R^{-\ell/2}$. The leading-order term can be further simplified as follows
\begin{equation}
\begin{aligned}
\label{03092020_1330}
\Bigl[ \mathscr{G}_{\mathfrak{s}}(x_{1},y;x_{2},-y) \Bigr]_{0} & = \Bigl[ G(x_{1},y;x_{2},-y) \Bigr]_{0} \, , \\
& = \frac{ \Delta \langle \sigma_{1} \rangle \Delta \langle \sigma_{2} \rangle }{ 4 } \biggl[ \mathcal{G}(\eta_{1},\tau;\eta_{2},-\tau) - \Upsilon(\chi_{1}) \Upsilon(\chi_{2}) \biggr] \, .
\end{aligned}
\end{equation}
The first identity follows by noticing that the excess part does not contribute at leading order, i.e., $\bigl[ \mathfrak{e}_{ab}(x_{j},y_{j}) \bigr]_{0}=0$. By using Mehler's decomposition of the joint passage probability density it is possible to write the scaling function $\mathcal{G}$ as a series of factorized terms. Leaving the details in Appendix \ref{appB}, we find
\begin{equation}
\label{03092020_1329}
\mathcal{G}(\eta_{1},\tau_{1};\eta_{2},\tau_{2}) = \Upsilon(\chi_{1})\Upsilon(\chi_{2}) + \sum_{n=1}^{\infty} \rho^{2n} \Upsilon_{2n+1}(\chi_{1}) \Upsilon_{2n+1}(\chi_{2}) \, ,
\end{equation}
where $\rho$ is the correlation coefficient given by (\ref{03092020_1250}) and $\Upsilon_{2n+1}(\chi_{j})$ are the functions defined by (\ref{03092020_1251}). By inserting (\ref{03092020_1329}) into (\ref{03092020_1330}) and focusing on identical spin fields, i.e., $\sigma_{1}=\sigma_{2}\equiv\sigma$, we find
\begin{equation}
\label{03092020_0941}
\Bigl[ \mathscr{G}_{\mathfrak{s}}(x_{1},y_{1};x_{2},y_{2}) \Bigr]_{0} = \frac{ (\Delta \langle \sigma \rangle)^{2} }{ 4 } \sum_{n=1}^{\infty} \rho^{2n} \Upsilon_{2n+1}(\chi_{1}) \Upsilon_{2n+1}(\chi_{2}) \, .
\end{equation}
The subleading correction at order $R^{-1/2}$ is formally given by
\begin{equation}
\begin{aligned}
\label{02062021_0915}
\Bigl[ \mathscr{G}_{\mathfrak{s}}(x_{1},y_{1};x_{2},y_{2}) \Bigr]_{1} & = \Bigl[ G(x_{1},y_{1};x_{2},y_{2}) \Bigr]_{1} + \Bigl[ \mathfrak{e}_{ab}(x_{1},y_{1}) \mathfrak{e}_{ab}(x_{2},y_{2}) \Bigr]_{1} \, ,
\end{aligned}
\end{equation}
however, the second addend vanishes because $\mathfrak{e}_{ab}$ is itself of order $R^{-1/2}$, therefore only the first addend contributes and one is left with
\begin{equation}
\begin{aligned}
\label{}
\Bigl[ \mathscr{G}_{\mathfrak{s}}(x_{1},y_{1};x_{2},y_{2}) \Bigr]_{1} & = \Bigl[ \langle \sigma_{1}(x_{1},y_{1}) \sigma_{2}(x_{2},y_{2}) \rangle_{B_{bab}} - \langle \sigma_{1}(x_{1},y_{1}) \rangle_{B_{bab}} \langle \sigma_{2}(x_{2},y_{2}) \rangle_{B_{bab}} \Bigr]_{1} \\
& =  \Bigl[ \langle \sigma_{1}(x_{1},y_{1}) \sigma_{2}(x_{2},y_{2}) \rangle_{B_{bab}} \Bigr]_{1} - \Bigl[ \langle \sigma_{1}(x_{1},y_{1}) \rangle_{B_{bab}} \Bigr]_{1} \Bigl[ \langle \sigma_{2}(x_{2},y_{2})
\rangle_{B_{bab}} \Bigr]_{0} \\
& - \Bigl[ \langle \sigma_{1}(x_{1},y_{1}) \rangle_{B_{bab}} \Bigr]_{0} \Bigl[ \langle \sigma_{2}(x_{2},y_{2})
\rangle_{B_{bab}} \Bigr]_{1} \, .
\end{aligned}
\end{equation}
The explicit expression is readily obtained
\begin{equation}
\begin{aligned}
\label{091507}
\Bigl[ \mathscr{G}_{\mathfrak{s}}(x_{1},y_{1};x_{2},y_{2}) \Bigr]_{1} & = \frac{\omega_{1}\Delta\langle\sigma_{1}\rangle\Delta\langle\sigma_{2}\rangle}{m} \int_{0}^{x_{2}}\textrm{d}u_{2} \Bigl[ P_{1}(x_{1},y_{1}) P_{1}(u_{2},y_{2}) - P_{2}(x_{1},y_{1};u_{2},y_{2}) \Bigr] \\
& + \frac{\omega_{2}\Delta\langle\sigma_{1}\rangle\Delta\langle\sigma_{2}\rangle}{m} \int_{0}^{x_{1}}\textrm{d}u_{1} \Bigl[ P_{1}(u_{1},y_{1}) P_{1}(x_{2},y_{2}) - P_{2}(u_{1},y_{1};x_{2},y_{2}) \Bigr] \, .
\end{aligned}
\end{equation}

The next task is the calculation of the integrals with respect to $x_{1},x_{2}$ appearing in (\ref{03092020_0940}). Focusing on the leading order contribution of the connected correlation function, we show in Appendix \ref{appC} that it is possible to compute the integrals of (\ref{03092020_0941}) for each term in the sum and, moreover, it is also possible to resum the resulting series in closed form. Regarding the first subleading order contribution, we observe that (\ref{091507}) vanishes when either $x_{1}$ or $x_{2}$ tend to $+ \infty$ and that its integral with respect to $x_{1},x_{2}$ also vanishes, i,e.,
\begin{equation}
\label{ }
\int_{0}^{\infty}\textrm{d}x_{1} \int_{0}^{\infty}\textrm{d}x_{2} \, \Bigl[ \mathscr{G}_{\mathfrak{s}}(x_{1},y_{1};x_{2},y_{2}) \Bigr]_{1} = 0 \, ,
\end{equation}
therefore the subleading correction to the connected correlation function ($\ell=1$) does not contribute to the integrals involved in (\ref{03092020_0904}).

We are now in the position to collect the contributions to $\widehat{S}(q)$ generated by the leading and first subleading expressions for the pair correlation function of the order parameter. In Appendix \ref{appC}, we show that the integration of (\ref{03092020_0941}) admits the following expression
\begin{equation}
\begin{aligned}
\label{03092020_1343}
\int_{0}^{\infty}\textrm{d} x_{1} \int_{0}^{\infty}\textrm{d} x_{2} \, \Bigl[ \mathscr{G}_{\mathfrak{s}}(x_{1},y;x_{2},-y) \Bigr]_{0} & =  (\Delta\langle\sigma\rangle)^{2} \frac{ R }{ 2m } E(\tau) \, ,
\end{aligned}
\end{equation}
with
\begin{equation}
\begin{aligned}
\label{23052021_1433}
E(\tau) & = \frac{2}{\pi}(1-\tau) \left( -2 + 3\sqrt{\tau} - 2\tau \right) + \left( 3 - 2\tau + 3\tau^{2} \right) \biggl[ \frac{1}{2} - \frac{2}{\pi} \tan^{-1}(\sqrt{\tau}) \biggr] \, .
\end{aligned}
\end{equation}
Coming back to the definition (\ref{03092020_0904}), it is implicit that $q$ is much larger than the lower momentum cutoff imposed by the system size; hence, $q \gg q_{\min} \sim R^{-1}$. Analogously, the wavenumber $q$ cannot be larger than the upper momentum cutoff $q_{\max} \sim 1/\xi_{\rm b} \propto m$ set by the inverse bulk correlation length, which plays the role of a microscopic scale. With this in mind, the parallel Fourier transform is thus computed for $q \gg 1/R$ and the following result is obtained
\begin{equation}
\label{03092020_1338}
\widehat{S}(q) \simeq \frac{ 1 }{ m q^{2} } \biggl[ 1 - \frac{32}{\sqrt{\pi}} \frac{1}{(qR)^{3/2}} + \dots \biggr] \, .
\end{equation}
The mathematical details involved in such a derivation are collected in Appendix \ref{appC}. Since corrections at order $\ell=1$ do not report in (\ref{03092020_0940}), the expression (\ref{03092020_1338}) for the structure factor, which is computed from terms at order $\ell=0$ and $\ell=1$, actually does not involve specificities of the bulk universality class encoded in the coefficients $\omega_{j}$. Details involving the bulk universality class play a role in the terms at order $\ell \geqslant 2$ and could play a role through further subleading corrections in the large-$R$ expansion of the interface structure factor. One of these terms at $\ell=2$ is originated by the term $\mathfrak{e}_{ab}(x_{1},y_{1}) \mathfrak{e}_{ab}(x_{2},y_{2})$, which is of order $R^{-1}$. In Appendix \ref{appC}, we show that such a term yields the following contribution to the interface structure factor
\begin{equation}
\label{02062021_0912_1}
\Delta \widehat{S}(q) = \digamma \frac{\sin(qR/2)}{m^{2}q} \, ;
\end{equation}
with an overall factor
\begin{equation}
\label{02062021_0913}
\digamma = \left( \frac{c_{0}}{\Delta\langle\sigma\rangle} + \frac{\mathfrak{b}}{\mathfrak{a}} \right)^{2}
\end{equation}
that depends on the bulk and boundary universality classes through the quantities $c_{0}/\Delta\langle\sigma\rangle$ and $\mathfrak{b}/\mathfrak{a}$. The latter are known for integrable bulk and boundary field theories, respectively. Quite interestingly, the structure of the interface -- encoded in the factor $A_{ab} \sim \sqrt{\digamma}$ -- affects the interface structure factor $\widehat{S}(q)$. The result is the contribution (\ref{02062021_0912_1}) which however localizes towards $q=0$ in the limit of large $qR$. This feature has been already reported in the exact investigation of the interface structure factor on the strip geometry \cite{DS_twopoint}. In that case the structure of the interface contributes to $\widehat{S}(q)$ through a completely analogous term, however, with a different value of the structure amplitude pertinent to the strip geometry, which reads $A_{ab} \propto c_{0}^{2}$. While the latter vanishes for the Ising model ($c_{0}=0$), $\digamma$ does vanish not because of the boundary data $\mathfrak{b}/\mathfrak{a}\neq0$.

The systematic calculation of these higher-order finite-size corrections to the spin-spin correlation function and their contribution to the structure factor certainly deserves further studies but goes far beyond the scope of the present analysis. On the other hand, the present analysis indicates the effect played by entropic repulsion of the interface from the wall. The entropic repulsion is already signaled by the quadratic term $\propto \chi^{2}$ in the passage probability $P_{1}(x,y)$. The latter affects the shape of the order parameter profile at any order in the large-$R$ expansion and is responsible for the additional term displayed in the square brackets of (\ref{03092020_1338}). Such a feature is indeed absent in the calculation on the strip geometry \cite{DS_twopoint} where no entropic effects come into play. By taking the limit of infinite system size (\ref{03092020_1338}) leads to
\begin{equation}
\label{}
S(q) = \lim_{R \rightarrow \infty} \widehat{S}(q) = \frac{1}{mq^{2}} \, .
\end{equation}
This is the typical Wertheim-Weeks divergence \cite{PR_NaturePhysics} at $q=0$ of the interface structure factor exhibited by fluid state theories theories in spatial dimensions $d \geqslant 3$ \cite{Wertheim_1976, Weeks_1977, BW_1985} and by the exact theory of phase separation in $d=2$ \cite{DS_twopoint}.

\section{Conclusions}
\label{sec_conclusions}
In this paper we showed how non perturbative techniques can be used in order to compute correlation functions for phase-separating systems on the half-plane. In particular, we studied the scaling limit of a ferromagnetic spin model at a first order transition close to a second order phase transition point. The system is defined on the half plane with boundary conditions enforcing the formation of a droplet which separates coexisting phases. More technically, the results of this paper follow as a combination of the exact field-theoretical formalism developed in \cite{DS_wetting} and \cite{DS_twopoint}.

Finite-size corrections arising from interface structure are interpreted according to a probabilistic picture. According to such a picture, the fluctuating droplet is interpreted as a Brownian excursion in one space dimension and whose probability distribution is found in closed form. The occurrence of long-range interfacial correlations mediated by a fluctuating droplet are thus established by closed-form expressions in real space including subleading corrections due to interface structure, which we also find. In general, exact results for correlations in real space emerge from general properties exhibited by bulk and boundary form factors at small momenta. The excellent agreement between the analytical results derived in this paper and Monte Carlo simulations for the Ising model will appear in a separate publication \cite{droplet_MC}. 

In the last part of the paper, we examined the long-range character of interfacial correlations in momentum space through the notion of interface structure factor. We showed how to extend traditional studies of the interface structure factor to a system defined on the semi-infinite space in two dimensions. After having isolated the interfacial degrees of freedom from the bulk ones, we have calculated the contribution of the leading order ($\ell=0$) and first subleading order ($\ell=1$) of the correlation function to the interface structure factor. We then showed how the entropic repulsion of the interface from the wall reflects into a specific term in the structure factor. In summary, the exact calculation given in this paper shows that the interface structure factor for the pinned droplet in two dimensions exhibits the long-wavelengths asymptotic behavior proportional to $1/q^{2}$. Although the latter feature is the typical result found within effective descriptions in terms of the capillary-wave theory for an interface which fluctuates in the \emph{bulk}, the calculation presented in this paper relies on an exact formalism built in terms of the \emph{fundamental} degrees of freedom of the near-critical system in contact with a boundary. In view of future directions, it would be interesting to characterize how the geometry affects correlations in a two-dimensional wedge. This paper indicates how to investigate the emergence of the elusive symmetry named \emph{wedge covariance} \cite{PR_2000, APW_2002, RP_2005, PR_2010, DS_wedge} at the level of correlation functions. The analysis of two-dimensional models could provide helpful insights towards the study of three-dimensional systems as well \cite{DSS_2020,DSS_2021}.

\section*{Acknowledgements}
A.~S is grateful to Gesualdo Delfino for his valuable comments and to Douglas B. Abraham for many interesting discussions and for collaborations on closely related topics.

\appendix
\section{Energy density correlations}
\label{appA}
In this appendix we show how to derive (\ref{37}). The saddle-point method, which has been used for the calculation of the one-point correlation function is used also for (\ref{32}). The function $U_{\pm}$ given by (\ref{03092020_1358}) is expanded at low energies and rapidities are rescaled as follows: $\theta_{j} \rightarrow \sqrt{2/(mR)}\theta_{j}$; thus, $U_{\pm}$ becomes
\begin{equation}
\begin{aligned}
\label{ }
Y_{\pm}(\theta_{1},\theta_{2},\theta_{3}) & = \exp\biggl[ - \frac{1-\tau_{1}}{2} \theta_{1}^{2} - \frac{\tau_{1}-\tau_{2}}{2} \theta_{2}^{2} - \frac{\tau_{2}+1}{2} \theta_{3}^{2} + \im \eta_{1} (\theta_{1} \mp \theta_{2}) + \im \eta_{2} (\theta_{2} - \theta_{3}) \biggr] \, ,
\end{aligned}
\end{equation}
up to a factor $\exp(-mR)$ which cancels the corresponding one from the partition function $\mathcal{Z}$ at the denominator of (\ref{32}). Hence, (\ref{32}) becomes
\begin{equation}
\begin{aligned}
\label{ }
\langle \varepsilon(x_{1},y_{1}) \varepsilon(x_{2},y_{2}) \rangle_{ B_{bab} }^{\rm CP} & = \frac{ 2 \left( F_{aba}^{\varepsilon}(\im \pi) \right)^{2}  }{ \pi^{5/2} mR } \int_{\mathbb{R}^{3}} \textrm{d}\theta_{1}\textrm{d}\theta_{2}\textrm{d}\theta_{3} \, \theta_{1}\theta_{3}  \biggl[ Y_{+}(\theta_{1},\theta_{2},\theta_{3}) - Y_{-}(\theta_{1},\theta_{2},\theta_{3}) \biggr] \, .
\end{aligned}
\end{equation}
The next task is to compute the Gaussian integrals in the above. To this end it is convenient to introduce the shorthand notation
\begin{equation}
\label{02092020_1057}
\Lbag \mathcal{O}(\theta_{1},\theta_{2},\theta_{3}) \Rbag_{\pm} \equiv \int_{\mathbb{R}^{3}}\prod_{j=1}^{3} \textrm{d}\theta_{j} \, \mathcal{O}(\theta_{1},\theta_{2},\theta_{3}) Y_{\pm}(\theta_{1},\theta_{2},\theta_{3}) \, ,
\end{equation}
therefore
\begin{equation}
\begin{aligned}
\label{03092020_1419}
\langle \varepsilon(x_{1},y_{1}) \varepsilon(x_{2},y_{2}) \rangle_{ B_{bab} }^{\rm CP} & = \frac{ 2 \left( F^{\varepsilon}(\im \pi) \right)^{2}  }{ \pi^{5/2} mR } \biggl[ \Lbag \theta_{1}\theta_{3} \Rbag_{+} - \Lbag \theta_{1}\theta_{3} \Rbag_{-} \biggr] \, .
\end{aligned}
\end{equation}
We observe that
\begin{equation}
\begin{aligned}
\label{ }
\Lbag \theta_{1}\theta_{3} \Rbag_{ \pm } & = \frac{ 4\pi^{5/2} \lambda^{2} \eta_{1}\eta_{2}}{(1-\tau_{1})(1+\tau_{2})} P_{\rm strip}(\pm x_{1},y_{1}; x_{2},y_{2}) \, ,
\end{aligned}
\end{equation}
where
\begin{equation}
\begin{aligned}
\label{14022021_1531}
P_{\rm strip}(x_{1},y_{1}; x_{2},y_{2}) & = \frac{1}{\pi\kappa_{1}\kappa_{2}\lambda^{2}\sqrt{1-\rho^{2}}} \exp\Biggl[ - \frac{ \chi_{1}^{2} + \chi_{2}^{2} -2\rho \chi_{1} \chi_{2} }{ 1-\rho^{2} } \Biggr] \\
& = \frac{2}{\kappa_{1}\kappa_{2}\lambda^{2}} \Pi_{2}(\sqrt{2}\chi_{1},\sqrt{2}\chi_{2} \vert \rho) \, ,
\end{aligned}
\end{equation}
is the passage probability on the strip. Equation (\ref{14022021_1531}) is the joint probability which characterizes a Brownian bridge in one space dimension \cite{DS_twopoint}. Thanks to (\ref{14022021_1531}), we can write
\begin{equation}
\begin{aligned}
\label{A001}
\Lbag \theta_{1}\theta_{3} \Rbag_{+} - \Lbag \theta_{1}\theta_{3} \Rbag_{-} & = \frac{ 4\pi^{5/2} \lambda^{2} \eta_{1}\eta_{2}}{(1-\tau_{1})(1+\tau_{2})} \biggl[ P_{\rm strip}(x_{1},y_{1}; x_{2},y_{2}) - P_{\rm strip}(-x_{1},y_{1};  x_{2},y_{2}) \biggr] \, , \\
& \equiv \pi^{5/2} \lambda^{2} P_{2}(x_{1},y_{1}; x_{2},y_{2})
\end{aligned}
\end{equation}
and (\ref{03092020_1419}) simplifies to
\begin{equation}
\begin{aligned}
\label{03092020_1420}
\langle \varepsilon(x_{1},y_{1}) \varepsilon(x_{2},y_{2}) \rangle_{ B_{bab} }^{\rm CP} & = \frac{2\lambda^{2}}{mR} \left( F^{\varepsilon}(\im \pi) \right)^{2} P_{2}(x_{1},y_{1}; x_{2},y_{2}) \, ,
\end{aligned}
\end{equation}
which is (\ref{37}) in the main body of the paper. Then, from (\ref{A001}), we obtain
\begin{equation}
\label{ }
P_{2}(x_{1},y_{1}; x_{2},y_{2}) = \frac{ 4\eta_{1}\eta_{2}}{(1-\tau_{1})(1+\tau_{2})} \biggl[ P_{\rm strip}(x_{1},y_{1}; x_{2},y_{2}) - P_{\rm strip}(-x_{1},y_{1};  x_{2},y_{2}) \biggr] \, .
\end{equation}
The expression (\ref{38}) for $P_{2}$ follows by expressing $P_{\rm strip}$ in terms of the bivariate Gaussian (\ref{appB01}). It has to be observed how the passage probability $P_{2}$ for the Brownian excursion is related to the passage probability of the Brownian bridge. It is indeed evident that (\ref{14022021_1531}) amounts the construction implied by the method of images.

\section{Parallel correlation function}
\label{appD}
In this appendix we show how to derive the analytic expression for the parallel correlation function given by (\ref{04092020_1629}). By setting $x_{1}=x_{2} \equiv x$ in the integral representation provided by (\ref{02092020_1426}), we have
\begin{equation}
\begin{aligned}
\label{0409_1620}
\mathcal{G}(\eta,\tau_{1};\eta,\tau_{2}) & = \int_{0}^{\infty}\textrm{d}u_{1} \int_{0}^{\infty}\textrm{d}u_{2} \, P_{2}(u_{1},y_{1};u_{2},y_{2}) \textrm{sign}(x-u_{1}) \textrm{sign}(x-u_{2}) \, ,
\end{aligned}
\end{equation}
by applying $\partial_{\eta}=\lambda\partial_{x}$ to both sides of (\ref{0409_1620})
\begin{equation}
\begin{aligned}
\label{n02}
\partial_{\eta}\mathcal{G}(\eta,\tau_{1};\eta,\tau_{2}) & = 2\lambda \int_{0}^{\infty}\textrm{d}u_{1} \int_{0}^{\infty}\textrm{d}u_{2} \, P_{2}(u_{1},y_{1};u_{2},y_{2}) \biggl[ \delta(x-u_{1}) \textrm{sign}(x-u_{2}) \\
& + \delta(x-u_{2}) \textrm{sign}(x-u_{1})  \biggr] \, , \\
& = 2\lambda \biggl[ \int_{0}^{\infty}\textrm{d}u_{2} \, P_{2}(x,y_{1};u_{2},y_{2}) \textrm{sign}(x-u_{2}) \\
& + \int_{0}^{\infty}\textrm{d}u_{1} \, P_{2}(u_{1},y_{1};x,y_{2}) \textrm{sign}(x-u_{1}) \biggr]
\end{aligned}
\end{equation}
and replacing the dummy variable in the second integral with $u_{2}$, we find
\begin{equation}
\begin{aligned}
\label{n02}
\partial_{\eta}\mathcal{G}(\eta,\tau_{1};\eta,\tau_{2}) & = 2\lambda \int_{0}^{\infty}\textrm{d}u_{2} \, \Bigl[ P_{2}(x,y_{1};u_{2},y_{2}) + P_{2}(u_{2},y_{1};x,y_{2}) \Bigr] \textrm{sign}(x-u_{2}) \, .
\end{aligned}
\end{equation}
In general, $P_{2}(x,y_{1};u_{2},y_{2}) \neq P_{2}(u_{2},y_{1};x,y_{2})$, however, if we take $\tau_{1}=\tau$ and $\tau_{2}=-\tau$, which is the case for the parallel correlation function we are interested in, we have the identity $P_{2}(x,y;u_{2},-y) = P_{2}(u_{2},y;x,-y)$, which implies
\begin{equation}
\begin{aligned}
\label{n02}
\partial_{\eta}\mathcal{G}(\eta,\tau;\eta,-\tau) & = 4\lambda \int_{0}^{\infty}\textrm{d}u \, P_{2}(x,y;u,-y) \textrm{sign}(x-u) \, .
\end{aligned}
\end{equation}
Thanks to (\ref{38}), the above reads
\begin{equation}
\begin{aligned}
\partial_{\eta}\mathcal{G}(\eta,\tau;\eta,-\tau) & = \frac{16\chi}{\kappa\rho} \Bigl[ I(\chi,\rho) - I(\chi,-\rho) \Bigl]
\end{aligned}
\end{equation}
with $\chi=\eta/\sqrt{1-\rho^{2}}$,
\begin{equation}
\begin{aligned}
I(\chi,\rho) & = \frac{\sqrt{1-\rho^{2}}}{2\pi} \Bigl[ \textrm{e}^{-\chi^{2}/(1-\rho^{2})} - 2 \textrm{e}^{-\chi^{2}/(1+\rho)} \Bigr] \\
& + \frac{\rho\chi}{2\sqrt{\pi}} \textrm{e}^{-\chi^{2}} \Bigl[ -1 + \textrm{erf}(\rho\chi/\sqrt{1-\rho^{2}}) + 2 \textrm{erf}(\sqrt{(1-\rho)/(1+\rho)}\chi) \Bigr] \, ,
\end{aligned}
\end{equation}
and $\rho = (1-\tau)/(1+\tau)$. A simple calculation entails
\begin{equation}
\begin{aligned}
\partial_{\eta}\mathcal{G}(\eta,\tau;\eta,-\tau) & = \frac{16\chi}{\pi\kappa\rho} \sqrt{1-\rho^{2}} \Bigl[ \textrm{e}^{-\frac{2\chi^{2}}{1-\rho}} - \textrm{e}^{-\frac{2\chi^{2}}{1+\rho}} \Bigl] \\
& + \frac{16\chi^{2}}{\sqrt{\pi}\kappa} \textrm{e}^{-\chi^{2}} \biggl[ -1 + \textrm{erf}\left(\sqrt{\frac{1-\rho}{1+\rho}}\chi\right) + \textrm{erf}\left(\sqrt{\frac{1+\rho}{1-\rho}}\chi\right) \biggl] \, .
\end{aligned}
\end{equation}
Equivalently
\begin{equation}
\begin{aligned}
\label{03092020_1434}
\partial_{\eta}\mathcal{G}(\eta,\tau;\eta,-\tau) & = \frac{16}{\sqrt{\pi}\kappa} \chi^{2} \textrm{e}^{-\chi^{2}} \biggl[ \textrm{erf}(\chi\sqrt{\tau}) + \textrm{erf}(\chi/\sqrt{\tau}) - 1 \biggr] - \frac{32}{\pi \kappa} \frac{\chi \sqrt{\tau}}{1-\tau} \textrm{e}^{-\chi^{2}} \biggl[ \textrm{e}^{-\chi^{2}\tau} - \textrm{e}^{-\chi^{2}/\tau} \biggr] \, .
\end{aligned}
\end{equation}
The parallel correlation function can be obtained by integrating with respect to $\eta$ as follows
\begin{equation}
\label{ }
\mathcal{G}(\eta,\tau;\eta,-\tau) = \mathcal{G}(0,\tau;0,-\tau) + \int_{0}^{\eta} \textrm{d}\eta^{\prime} \, \partial_{\eta^{\prime}}\mathcal{G}(\eta^{\prime},\tau;\eta^{\prime},-\tau)
\end{equation}
with $\mathcal{G}(0,\tau;0,-\tau)=1$ from (\ref{0409_1620}) and the normalization condition for $P_{2}$.

In order to get a quantitative understanding of (\ref{03092020_1434}), we provide a more explicit representation valid in the asymptotic regime $\tau \rightarrow 0$. For small $\tau$ it is possible to adopt the Taylor expansion
\begin{equation}
\begin{aligned}
\label{n02}
\partial_{\eta}\mathcal{G}(\eta,\tau;\eta,-\tau) & = \frac{32}{\pi} \eta \left( \eta^{2}-1 \right) \textrm{e}^{-\eta^{2}} \tau^{1/2} - \frac{32}{3\pi} \eta \left( 3-3\eta^{2}+\eta^{4} \right) \textrm{e}^{-\eta^{2}} \tau^{3/2} + \Os(\tau^{5/2}) \, .
\end{aligned}
\end{equation}
Integrating back with respect to $\eta$ and imposing the boundary condition at $\eta=0$, $\mathcal{G}(0,\tau;0,-\tau)=1$, we readily find
\begin{equation}
\begin{aligned}
\label{04092020_1353}
\mathcal{G}(\eta,\tau;\eta,-\tau) & = 1 - \frac{16}{\pi} \eta^{2} \textrm{e}^{-\eta^{2}} \tau^{1/2} - \frac{16}{3\pi} \eta^{2} \left( 1 - \eta^{2} \right) \textrm{e}^{-\eta^{2}} \tau^{3/2} + \Os(\tau^{5/2}) \, .
\end{aligned}
\end{equation}
The expression (\ref{04092020_1629}) follows accordingly.

\section{Mehler's decomposition of the spin-spin correlator}
\label{appB}
The bivariate Gaussian distribution is defined by
\begin{equation}
\label{appB01}
\Pi_{2}(x_{1},x_{2} \vert \rho) = \frac{1}{2\pi \sqrt{1-\rho^{2}}} \exp\biggl[ - \frac{ x_{1}^{2}+x_{2}^{2}-2\rho x_{1}x_{2} }{ 2(1-\rho^{2}) } \biggr] \, ,
\end{equation}
with $\rho$ the correlation coefficient. The expression (\ref{appB01}) satisfies the properties
\begin{equation}
\begin{aligned}
\label{appB02}
\int_{\mathbb{R}} \textrm{d}x_{2} \, \Pi_{2}(x_{1},x_{2} \vert \rho) & = \frac{1}{\sqrt{2\pi}}\textrm{e}^{-x_{1}^{2}/2} \equiv \Pi_{1}(x_{1}) \, , \\
\int_{\mathbb{R}^{2}} \textrm{d}x_{1}\textrm{d}x_{2} \, \Pi_{2}(x_{1},x_{2} \vert \rho) & = 1 \, .
\end{aligned}
\end{equation}
The Gaussian bivariate can be expressed as an infinite series of factorized products containing the factors $\Pi_{1}(x_{1})\Pi_{1}(x_{2})$. This is the content of Mehler \cite{Mehler} theorem, which establishes the following identity
\begin{equation}
\begin{aligned}
\label{appB03}
\Pi_{2}(x_{1},x_{2} \vert \rho) & = \sum_{\ell=0}^{\infty} \frac{(\rho/2)^{\ell}}{\ell!} H_{\ell}(x_{1}/\sqrt{2}) H_{\ell}(x_{2}/\sqrt{2}) \Pi_{1}(x_{1})\Pi_{1}(x_{2}) \, ,
\end{aligned}
\end{equation}
with $H_{\ell}$ Hermite polynomials. For $\rho=0$ the random variables $x_{1}$ and $x_{2}$ are uncorrelated and the corresponding joint distribution factorizes, i.e., $\Pi_{2}(x_{1},x_{2} \vert \rho) = \Pi_{1}(x_{1})\Pi_{1}(x_{2})$, as expected.

\begin{figure}[htbp]
\centering
\includegraphics[width=11.5cm]{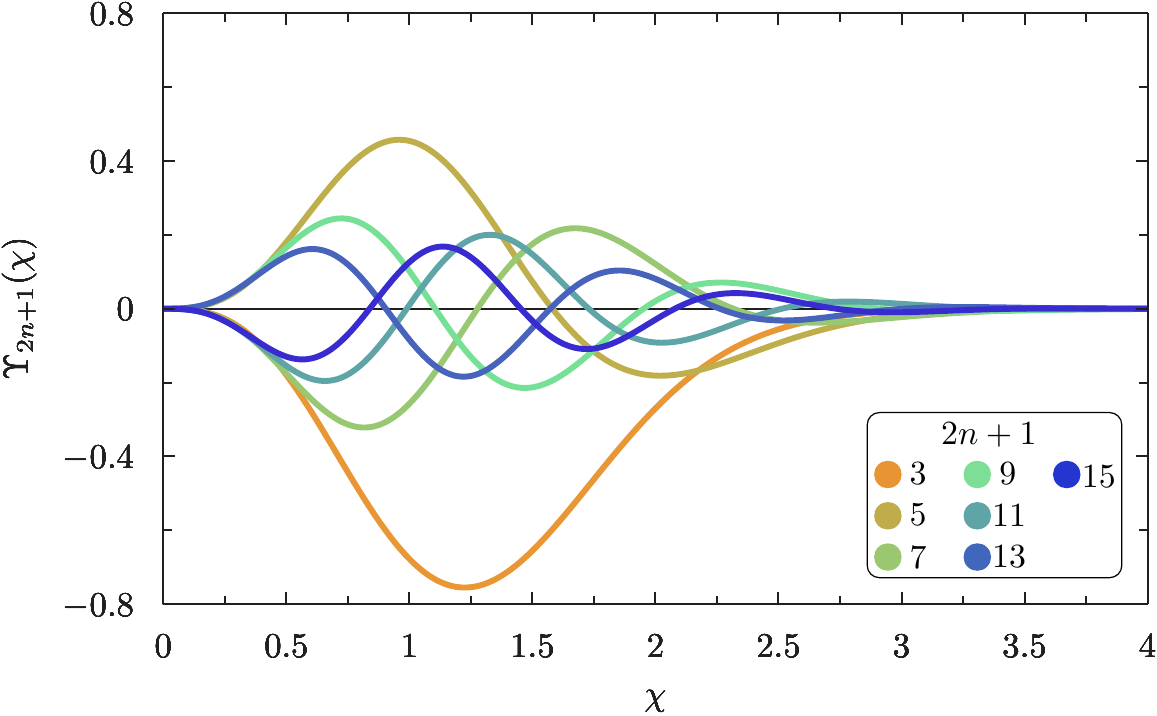}
\caption{The functions $\Upsilon_{2n+1}(\chi)$ for the values of $2n+1$ indicated in the inset.}
\label{fig_upsilonfunctions}
\end{figure}

By plugging (\ref{appB03}) into the joint passage probability density for the droplet-shaped interface (\ref{38}), we obtain
\begin{equation}
\label{appB04}
P_{2}(x_{1},y_{1}; x_{2},y_{2}) = \frac{ 8 \chi_{1} \chi_{2} }{ \pi \rho \kappa_{1} \kappa_{2} \lambda^{2} } \sum_{n=0}^{\infty} \frac{ (\rho/2)^{2n+1} }{ (2n+1)! } H_{2n+1}(\chi_{1}) H_{2n+1}(\chi_{2}) \textrm{e}^{ - \chi_{1}^{2} - \chi_{2}^{2} } \, .
\end{equation}
Thanks to (\ref{appB04}) the connected spin-spin correlation function (\ref{02092020_1426}) admits the series representation
\begin{equation}
\label{appB05}
\mathcal{G}(\eta_{1},\tau_{1};\eta_{2},\tau_{2}) = \sum_{n=0}^{\infty} \rho^{2n} \Upsilon_{2n+1}(\chi_{1}) \Upsilon_{2n+1}(\chi_{2}) \, ,
\end{equation}
where $\Upsilon_{2n+1}(\chi)$ are the functions
\begin{equation}
\label{03092020_1251}
\Upsilon_{2n+1}(\chi) = \frac{1}{2^{n-1}\sqrt{(2n+1)!\pi}} \int_{0}^{\infty}\textrm{d}v \, v H_{2n+1}(v) \textrm{e}^{-v^{2}} \textrm{sign}(\chi-v) \, ,
\end{equation}
the first few of them are plotted in Fig.~\ref{fig_upsilonfunctions}.

By using the definition of Hermite polynomials,
\begin{equation}
\label{ }
H_{n}(x) = (-1)^{n} \textrm{e}^{x^{2}} \frac{\rd^{n}}{\rd x^{n}} \textrm{e}^{-x^{2}} \, ,
\end{equation}
and integrating by parts in (\ref{03092020_1251}), we find the following representation of the functions $\Upsilon_{2n+1}(\chi)$ in terms of Hermite polynomials
\begin{equation}
\label{05092020_1129}
\Upsilon_{2n+1}(\chi) = - \frac{1}{2^{n-2}\sqrt{(2n+1)!\pi}} \biggl[ H_{2n-1}(\chi) + \chi H_{2n}(\chi) \biggr] \textrm{e}^{-\chi^{2}}  \, , \qquad n \geqslant 1 \, .
\end{equation}
It thus follows that $\Upsilon_{2n+1}(\chi)$ for $n\geqslant1$ satisfies $\Upsilon_{2n+1}(0) = \Upsilon_{2n+1}(\chi\rightarrow+\infty)=0$. Moreover, $\Upsilon_{2n+1}(\chi)$ with $n\geqslant1$ are localized functions in the sense that their integral is finite, while $\Upsilon_{1}(\chi)=\Upsilon(\chi)$ corresponds to an extended profile because it interpolates between $-1$ and $+1$; see the green curve in Fig.~\ref{fig_drops}.

\section{Calculation of $\widehat{S}(q)$}
\label{appC}
In this section, we complete the main steps involved in the calculation of the interface structure factor. We begin by computing the integral with respect to $x_{1}$ and $x_{2}$ of the function $[\mathscr{G}_{\mathfrak{s}}]_{0}$ appearing in (\ref{03092020_0940}) and defined by (\ref{03092020_0941}). To this end it is convenient to introduce the shorthand notation
\begin{equation}
\label{23052021_1400}
\llbracket \Upsilon_{2n+1} \rrbracket \equiv \int_{0}^{\infty}\textrm{d}\chi \, \Upsilon_{2n+1}(\chi) \, , \qquad n \geqslant1 \, .
\end{equation}
By plugging (\ref{05092020_1129}) into (\ref{23052021_1400}) and integrating by parts, we find
\begin{equation}
\begin{aligned}
\label{}
\llbracket \Upsilon_{2n+1} \rrbracket & = - \frac{H_{2n-2}(0)}{2^{n-3}\sqrt{(2n+1)!\pi}} \\
& = (-1)^{n} \frac{ 4 (2n-3)!! }{ \sqrt{(2n+1)!\pi} } \, ;
\end{aligned}
\end{equation}
in the last line we used Hermite numbers $H_{2n}(0)=(-2)^{n}(2n-1)!!$. The integral with respect to $x_{1}$ and $x_{2}$ of the connected correlation function reads
\begin{equation}
\begin{aligned}
\label{03092020_1343}
\int_{0}^{\infty}\textrm{d} x_{1} \int_{0}^{\infty}\textrm{d} x_{2} \, \bigl[ \mathscr{G}_{\mathfrak{s}}(x_{1},y;x_{2},-y) \bigr]_{0} & = (\Delta\langle\sigma\rangle)^{2}  \frac{ \kappa^{2} \lambda^{2}  }{ 4 } \sum_{n=1}^{\infty} \rho^{2n} \llbracket \Upsilon_{2n+1} \rrbracket^{2} \\
& = (\Delta\langle\sigma\rangle)^{2}  \frac{ 2\kappa^{2} R }{ \pi m } \sum_{n=1}^{\infty} \frac{ [(2n-3)!!]^{2} }{ (2n+1)! } \rho^{2n} \, .
\end{aligned}
\end{equation}
The series appearing in the right hand side of (\ref{03092020_1343}) can be re-summed thanks to the identity
\begin{equation}
\label{03092020_1344}
\sum_{n=1}^{\infty} \frac{ [(2n-3)!!]^{2} }{ (2n+1)! } \rho^{2n} = \left( \frac{\rho}{2} + \frac{1}{4\rho} \right) \sin^{-1}\rho + \frac{3}{4} \sqrt{1-\rho^{2}} - 1 \equiv H(\rho) \, .
\end{equation}
Turning to the interface structure factor, we have
\begin{equation}
\begin{aligned}
\label{120208}
\widehat{S}(q) & = \frac{ 1 }{ 2 \left( \Delta \langle \sigma \rangle \right)^{2} } \int_{-R/2}^{R/2}\textrm{d}y \, \textrm{e}^{\im q y } \int_{0}^{\infty}\textrm{d}x_{1} \int_{0}^{\infty}\textrm{d}x_{2} \, \bigl[ \mathscr{G}_{\mathfrak{s}}(x_{1},y;x_{2},-y) \bigr]_{0} \, , \\
& = \frac{ R^{2} }{ \pi m } \int_{0}^{1}\textrm{d}\tau \, \left( 1-\tau^{2} \right) H\left( \frac{1-|\tau|}{1+|\tau|} \right) \cos( Q \tau ) \, , \\
& = \frac{ R^{2} }{ 4 m } \int_{0}^{1}\textrm{d}\tau \, E(\tau) \cos( Q \tau ) \, , \\
\end{aligned}
\end{equation}
where
\begin{equation}
\label{ }
E(\tau) = \frac{4}{\pi} (1-\tau^{2}) H\left( \frac{1-|\tau|}{1+|\tau|} \right) \, ,
\end{equation}
is the function (\ref{23052021_1433}) whose explicit expression can be derived from (\ref{03092020_1344}) thanks to the identity
\begin{equation}
\label{ }
\sin^{-1}\left(\frac{1-\tau}{1+\tau}\right) = \frac{\pi}{2} - 2 \tan^{-1}(\sqrt{\tau}) \, .
\end{equation}
In the second line of (\ref{120208}), we rescaled the integration variable $y=(R/2)\tau$ and the rescaled wavenumber $Q=qR/2$ has been adopted. We also recall that for $y<0$ we have to revert the order of the two spins, this can be achieved by replacing $\tau$ with $|\tau|$. For large $Q$ the integral in (\ref{120208}) yields the asymptotic expansion
\begin{equation}
\begin{aligned}
\label{12062021_1654}
\widehat{S}(q) & = \frac{ R^{2} }{ 4m } \biggl[ \frac{1}{Q^{2}} - 8\sqrt{\frac{2}{\pi}} \frac{1}{Q^{7/2}} + \dots \biggr]
\end{aligned}
\end{equation}
which corresponds to (\ref{03092020_1338}) in the main body of the paper. The result (\ref{12062021_1654}) is proved in Appendix \ref{appendix_E}.

To conclude, we compute the contribution of the excess part to the interface structure factor. Denoting such a contribution with $\Delta \widehat{S}(q)$, we have
\begin{equation}
\begin{aligned}
\label{02062021_0833}
\Delta \widehat{S}(q) & = \frac{ 1 }{ 2 \left( \Delta \langle \sigma \rangle \right)^{2} } \int_{-R/2}^{R/2}\textrm{d}y \, \textrm{e}^{\im q y } \int_{0}^{\infty}\textrm{d}x_{1} \int_{0}^{\infty}\textrm{d}x_{2} \, \mathfrak{e}_{ab}(x_{1},y) \mathfrak{e}_{ab}(x_{2},-y)
\end{aligned}
\end{equation}
with the excess part given by $\mathfrak{e}_{ab}(x_{j},y_{j})=-A_{ab} P_{1}(x_{1},y_{1})$, up to subdominant corrections; see (\ref{02062021_0830}). Since $P_{1}$ is normalized, we have
\begin{equation}
\label{ }
\int_{0}^{\infty}\textrm{d}x_{1} \mathfrak{e}_{ab}(x_{1},y) = - A_{ab} 
\end{equation}
with the structure amplitude $A_{ab}$ given by (\ref{01092020_2047}). Therefore (\ref{02062021_0833}) becomes
\begin{equation}
\label{02062021_0912}
\Delta \widehat{S}(q) = \frac{ A_{ab}^{2} }{ 2 \left( \Delta \langle \sigma \rangle \right)^{2} } \int_{-R/2}^{R/2}\textrm{d}y \, \textrm{e}^{\im q y }
\end{equation}
or equivalently, by using (\ref{01092020_2047}), we can write
\begin{equation}
\Delta \widehat{S}(q) = \digamma \frac{\sin(qR/2)}{m^{2}q}
\end{equation}
with the overall factor $\digamma$ given by (\ref{02062021_0913}) in the main body of the paper.

\section{Special integrals}
\label{appendix_E}
In this appendix we show how to derive (\ref{12062021_1654}). We begin by considering the function $E(\tau)$ which appears in the integrand of (\ref{120208}) and whose expression -- given by (\ref{23052021_1433}) -- can be written as follows
\begin{equation}
\label{12062022_1713}
E(\tau) = E_{1}(\tau) + E_{2}(\tau) \, ,
\end{equation}
where
\begin{equation}
\begin{aligned}
\label{}
E_{1}(\tau) & = \frac{2}{\pi}(1-\tau) \left( -2 + 3\sqrt{\tau} - 2\tau \right) + \frac{1}{2} \left( 3 - 2\tau + 3\tau^{2} \right) \\
E_{2}(\tau) & = \left( - \frac{6}{\pi} + \frac{4}{\pi} \tau - \frac{6}{\pi} \tau^{2} \right) \tan^{-1}(\sqrt{\tau}) \, .
\end{aligned}
\end{equation}
The decomposition (\ref{12062022_1713}) entails $\widehat{S}(q)=\widehat{S}_{1}(q)+\widehat{S}_{2}(q)$, where
\begin{equation}
\begin{aligned}
\widehat{S}_{j}(q) & = \frac{ R^{2} }{ 4 m } \int_{0}^{1}\textrm{d}\tau \, E_{j}(\tau) \cos( Q \tau ) \, , \\
\end{aligned}
\end{equation}
for $j=1,2$. The calculation of $\widehat{S}_{1}(q)$ is immediate and the result is
\begin{equation}
\begin{aligned}
\widehat{S}_{1}(q) & = \frac{ R^{2} }{ 4 m } \biggl[ \frac{1}{Q^{2}} + \left( 2 - \frac{1}{\pi} \right) \frac{\cos Q}{Q^{2}} + 2 \frac{\sin Q}{Q} -  \left( 3 + \frac{8}{\pi} \right) \frac{\sin Q}{Q^{3}} + \frac{ 9 \mathcal{c}(Q) -6 Q \mathcal{s}(Q) }{\sqrt{2\pi}Q^{5/2}} \biggr] \, ,
\end{aligned}
\end{equation}
where $\mathcal{s}(Q)$ and $\mathcal{c}(Q)$ are the functions
\begin{equation}
\begin{aligned}
\label{12062021_1807}
\mathcal{s}(Q) & \equiv S(\sqrt{2Q/\pi}) \\
\mathcal{c}(Q) & \equiv C(\sqrt{2Q/\pi}) \, ,
\end{aligned}
\end{equation}
$S$ and $C$ are Fresnel integrals, whose definition is recalled for completeness
\begin{equation}
\begin{aligned}
\label{}
S(z) & = \int_{0}^{z}\textrm{d}u \, \sin\left(\frac{\pi}{2}u^{2}\right) \\
C(z) & = \int_{0}^{z}\textrm{d}u \, \cos\left(\frac{\pi}{2}u^{2}\right) \, .
\end{aligned}
\end{equation}

The calculation of $\widehat{S}_{2}(q)$ is less immediate. In order to proceed with its calculation, we introduce the following auxiliary functions
\begin{equation}
\begin{aligned}
\mathcal{V}_{n}(Q) & = \int_{0}^{1}\textrm{d}\tau \, \tau^{n} \tan^{-1}(\sqrt{\tau}) \cos(Q\tau) \, ,
\end{aligned}
\end{equation}
whose calculation will be presented shortly. Thanks to the above definition, we have
\begin{equation}
\begin{aligned}
\widehat{S}_{2}(q) & = - \frac{6}{\pi} \mathcal{V}_{0}(Q) + \frac{4}{\pi} \mathcal{V}_{1}(Q) - \frac{6}{\pi} \mathcal{V}_{2}(Q) \, .
\end{aligned}
\end{equation}
In order to simplify the calculation of $\mathcal{V}_{n}(Q)$ it is convenient to write $\cos(Q\tau)$ in exponential form; thus, we consider
\begin{equation}
\begin{aligned}
\mathcal{W}_{n}(Q) & = \int_{0}^{1}\textrm{d}\tau \, \tau^{n} \tan^{-1}(\sqrt{\tau}) \textrm{e}^{\im Q \tau} \, ,
\end{aligned}
\end{equation}
which implies that
\begin{equation}
\label{ }
\mathcal{V}_{n}(Q) = \frac{1}{2} \bigl[ \mathcal{W}_{n}(Q) + \mathcal{W}_{n}(-Q) \bigr] \, .
\end{equation}
The advantage is that $\mathcal{W}_{n}(Q)$ can be computed by employing Feynman's trick,
 meaning that
\begin{equation}
\mathcal{W}_{n}(Q) = \left( - \im \partial_{Q} \right)^{n} \mathcal{W}_{0}(Q) \, .
\end{equation}
As a result, the only integral we have to perform is the one which defines the function $\mathcal{W}_{0}(Q)$.

The calculation $\mathcal{W}_{0}(Q)$ can be carried out by writing the arctangent in terms of the integral representation
\begin{equation}
\label{ }
\tan^{-1}(\sqrt{\tau}) = \int_{0}^{\sqrt{\tau}} \frac{\textrm{d}x }{1+x^{2}} \, , 
\end{equation}
thus,
\begin{equation}
\begin{aligned}
\mathcal{W}_{0}(Q) & = \int_{0}^{1}\textrm{d}\tau \, \textrm{e}^{\im Q \tau} \int_{0}^{\infty} \frac{\textrm{d}x}{1+x^{2}} \theta(\sqrt{\tau}-x) \, ,
\end{aligned}
\end{equation}
where $\theta$ is Heaviside's theta function. By exchanging the order of integrations, and performing the integral with respect to $\tau$, we find
\begin{equation}
\begin{aligned}
\mathcal{W}_{0}(Q) & = \int_{0}^{\infty} \frac{\textrm{d}x}{1+x^{2}} \int_{0}^{1}\textrm{d}\tau \, \textrm{e}^{\im Q \tau}  \theta(\sqrt{\tau}-x) \\
& = \frac{\im}{Q}  \int_{0}^{\infty} \frac{\textrm{d}x}{1+x^{2}} \biggl[ \textrm{e}^{\im Q x^{2}} - \textrm{e}^{\im Q} \biggr] \theta(1-x) \\
& = \frac{\im}{Q} \int_{0}^{1} \frac{\textrm{d}x}{1+x^{2}} \biggl[ \textrm{e}^{\im Q x^{2}} - \textrm{e}^{\im Q} \biggr] \, ,
\end{aligned}
\end{equation}
hence
\begin{equation}
\mathcal{W}_{0}(Q) = - \frac{\im \pi }{4Q} \textrm{e}^{\im Q} + \frac{\im}{Q} \int_{0}^{1} \textrm{d}x \, \frac{\textrm{e}^{\im Q x^{2}}}{1+x^{2}} \, .
\end{equation}
The integral appearing in the second term can be calculated by recalling the following identity
\begin{equation}
\label{ }
\int_{0}^{1} \textrm{d}x \, \frac{\textrm{e}^{\im Q x^{2}}}{1+x^{2}} = \frac{\pi}{4} \textrm{e}^{-\im Q} \biggl[ 1 - \textrm{erf}^{2}\left( \sqrt{-\im Q}\right) \biggr] \, ,
\end{equation}
a further simplification occurs by bringing in Fresnel integrals thanks to the property
\begin{equation}
\label{ }
C(z) + \im S(z) = \frac{1+\im}{2} \textrm{erf}\left(\frac{\sqrt{\pi}}{2}(1-\im)z\right) \, ,
\end{equation}
from which it follows that
\begin{equation}
\label{ }
\textrm{erf}\left( \sqrt{-\im Q}\right) = \frac{2}{1+\im} \biggl[ \mathcal{c}(Q) + \im \mathcal{s}(Q) \biggr] \, ,
\end{equation}
where $\mathcal{c}(Q)$ and $\mathcal{s}(Q)$ are the functions defined by (\ref{12062021_1807}). The above results allow us to write
\begin{equation}
\begin{aligned}
\mathcal{W}_{0}(Q) & = \frac{\pi}{2} \frac{\sin Q}{Q}  - \frac{\pi}{2Q} \textrm{e}^{-\im Q} \biggl[ \mathcal{c}(Q) + \im \mathcal{s}(Q) \biggr]^{2} \, .
\end{aligned}
\end{equation}

The assemblage of the final result for $\widehat{S}(q)$ is now an elementary (but rather tedious) exercise which gives
\begin{equation}
\begin{aligned}
\label{12062021_2026}
\widehat{S}(Q) & = \frac{1}{mq^{2}} \biggl[ 1 - 4 \left( 1 + \frac{1}{\pi} \right) \cos Q + \left( 3 - \frac{8}{\pi} - 4Q^{2} \right) \frac{\sin Q}{Q} \\
& + 4 \sqrt{\frac{2}{\pi Q}} \left( 3\mathcal{c}(Q) - 2Q\mathcal{s}(Q) \right) \\
& + 4 \mathcal{s}(Q) \mathcal{c}(Q)  \left( 4\cos Q + 4 Q\sin Q - 3 \frac{\sin Q}{Q} \right) \\
& + \left( \mathcal{c}^{2}(Q) - \mathcal{s}^{2}(Q) \right) \left( 8 Q \cos Q - 8 \sin Q - 6 \frac{\cos Q}{Q} \right) \biggr] \, .
\end{aligned}
\end{equation}
By using the asymptotic expressions for the Fresnel integrals at large arguments, we find
\begin{equation}
\begin{aligned}
\mathcal{s}(Q) & = \frac{1}{2} + \frac{1}{\sqrt{2\pi Q}} \biggl[ - \cos Q - \frac{\sin Q}{2Q} + \frac{3}{4} \frac{\cos Q}{Q^{2}} \biggr] + O(Q^{-7/2}) \\
\mathcal{c}(Q) & = \frac{1}{2} + \frac{1}{\sqrt{2\pi Q}} \biggl[ \sin Q - \frac{\cos Q}{2Q} - \frac{3}{4} \frac{\sin Q}{Q^{2}} \biggr] + O(Q^{-7/2}) \, ,
\end{aligned}
\end{equation}
once we plug the above into (\ref{12062021_2026}), we obtain
\begin{equation}
\widehat{S}(Q) = \frac{1}{mq^{2}} \biggl[ 1 - 8 \sqrt{\frac{2}{\pi}} \frac{1}{Q^{3/2}} + O(Q^{-2}) \biggr] \, ;
\end{equation}
which proves the result given by (\ref{12062021_1654}).


\bibliographystyle{unsrt}
\bibliography{bibliography}{}

\end{document}